\documentclass[structabstract]{aa}

\usepackage{graphicx}

\usepackage{txfonts}

\usepackage{natbib}
\bibpunct{(}{)}{;}{a}{}{,} % to follow the A&A style
\begin{document}

% http://www.eso.org/sci/observing/policies/publications.html
\title{The effect of stellar limb darkening values on the accuracy of the 
planet radii derived from photometric transit observations}

%\subtitle{Research Note} 

\author{Sz.~Csizmadia$^1$, 
        Th.~Pasternacki$^1$,
	C.~Dreyer$^{1,2}$,
	J.~Cabrera$^1$,
	A.~Erikson$^1$,
	H.~Rauer$^{1,2}$
	}

\institute{
Institute of Planetary Research, German Aerospace Center, Rutherfordstrasse 2, 12489 Berlin, Germany, \email{szilard.csizmadia@dlr.de}
 \and
Zentrum f\"ur Astronomie und Astrophysik, TU Berlin, Hardenbergstrasse 36, 10623 Berlin, Germany
}

\date{Received 2012 Jun 25; Accepted 2012 Sep 6}

\abstract
   {The radius of an exoplanet is one of its most important parameters.  
    Studies of planetary interiors and their evolution require 1\% precision   in 
    the radius determination. Transiting exoplanets offer a unique oppurtunity 
    to measure the radius of exoplanets in stellar units. These radius 
    measurements and their precision are strongly affected by our knowledge of 
    limb darkening.}
   {We study how the precision of the exoplanet radius determination is affected
    by our present knowledge of limb darkening in two cases: when we fix the
    limb darkening coefficients and when we adjust them. We also investigate
    the effects of spots in one-colour photometry.}
   {We study the effect of limb darkening on the planetary radius determination
   both via analytical expressions and by numerical experiments. We also compare
   some of the existing limb darkening tables. When stellar spots affect the
   fit, we replace the limb darkening coefficients, calculated for the
   unspotted cases, with effective limb darkening coefficients to describe the
   effect of the spots.}
   {There are two important cases. (1) When one fixes the limb darkening values
   according to some theoretical predictions, the inconsistencies of the tables
   do not allow us to reach accuracy in the planetary radius of better than 1-10\%
   (depending on the impact parameter) if the host star's surface effective
   temperature is higher than $5000$K. Below $5000$K the radius ratio
   determination may contain even 20\% error. (2) When one allows adjustment of the
   limb darkening coefficients, the $a/R_s$ ratio, the planet-to-stellar radius
   ratio, and the impact parameter can be
   determined with sufficient accuracy ($<1$\%), if the signal-to-noise ratio is
   high enough. However, the presence of stellar spots
   and faculae can destroy the agreement between the limb darkening tables and
   the fitted limb darkening coefficients, but this does not affect the
   precision of the planet radius determination. We also find that it is
   necessary to fit the contamination factor, too.}
   {We conclude that the present inconsistencies of theoretical stellar limb 
   darkening tables suggests one should not fix the limb darkening coefficients. When
   one allows them to be adjusted, then the planet radius, impact parameter, and the
   $a/R_{s}$ can be obtained with the required precision.}

%% traditional abstract format  
%\abstract{}
\keywords{stars: planetary systems - techniques: photometry }

\titlerunning{Transit parameters, stellar spots and limb darkening}
\authorrunning{Csizmadia et al.}

\maketitle

%
%____________________________________________________________________________
\section{Introduction}
\label{sec:introduction}

The depth of an exoplanetary transit is primarily determined by the ratio of the
radius of the transiting exoplanet to the host star. Limb darkening is a
second-order, but key effect for determining the exact radius of the planet,
because limb darkening not only modifies the shape of the transit light curve,
but it also significantly affects the true transit depth. There are other
third-order effects (e.g. night-side radiation, exorings, gravity darkening, 
etc.), but they are not studied here.

Transit parameters are determined by fitting a model to the observed light
curve  data points. Typically the number of free parameters is around seven (e.g. epoch, period, duration of the transit, the impact
parameter, radius ratio of the planet and the host star, and two limb darkening
coefficients). If we knew a priori the exact values of the limb darkening
coefficients, then the number of free parameters would be significantly
reduced by fixing the limb darkening coefficients. This would mean not only that the
dimensions of this optimization problem are reduced, but also that the occasional degeneracy of the
fitting procedure reported by several authors would disappear (e.g. Brown et al.
2001; Deleuil et al. 2008; Csizmadia et al. 2011). 

Values of the limb darkening coefficients have been published by several authors for
many  photometric passbands as a function of the stellar effective temperature,
metallicity,  $\log g$, and turbulent velocity in the stellar atmosphere. Such
tables can  be found in van Hamme (1993), Diaz-Cordoves et al. (1995), Claret et
al. (1995), Claret  (2000, 2004), Claret \& Hauschildt (2003), Barban (2003), 
Sing (2010), Howarth (2011), and Claret \& Bloemen (2011). For  earlier tables see
the summary of van Hamme (1993).

Some investigators of transit light curves have found good agreement  between
theoretical predictions and fitted limb darkening coefficients  (e.g. in
CoRoT-8, Bord\'e et al 2010, or in CoRoT-11 Gandolfi et al.  2010); however,
sometimes larger differences were found (e.g. at  $2\sigma$ level in CoRoT-13,
Cabrera et al. 2010) or the fit did not  constrain the values of the limb
darkening coefficients at all (e.g. in  CoRoT-12 Gillon 2010). Complete
disagreement can often be found between  the theoretically predicted and the
observed limb darkening  coefficients, even where the quality of the light curve
is extremely  good (e.g. the analysis of the HST light curve of HD 209~458 by 
Claret (2009), re-analysis of Kepler-5b by  Kipping \& Bakos (2011)); or else, the ground-based photometry of WASP-13  
by (Barros et al. 2012), who also point out that these discrepancies are not caused
by the photometric signal-to-noise ratio (S/N), and the quality of the light curves are
sufficient to say that  aforementioned discrepancies are real. Claret (2009)
concludes that uncertainties in the stellar parameters (effective temperature,
surface  gravity ($\log g$), metallicity) are also not responsible for the 
discrepancies.

Neilson \& Lester (2011) point out that the used plan-parallel
theory might be replaced by spherical symmetry for calculating limb
darkening coefficients, which may provide better predictions for limb darkening. 
The 3D stellar atmosphere models also offer an oppurtunity to improve the predictions.
However, these very new models have not been tested against full and multicolour light
curves yet (Hayek et al. 2012).

Since limb darkening changes true transit depths (see Appendix), the question
arises: do these insonsistencies affect the accuracy of the transit parameter
determination? As we show, the answer is yes. In addition, to distinguish
between different planetary models and to study them in detail we require
$\pm1$\% precision in planetary radii below five Earth masses (Wagner et al 2011;
F. Sohl, priv. comm.), a level that cannot be reached in all temperature regions
when fixing the limb darkening coefficients for the transit fit.

The aim of this paper is not only to take a step toward better understanding 
what can cause inconsistent observed limb darkening coefficients. We also want to make certain 
 that such inconsistencies between observed and predicted limb
darkening coefficients do not affect the determination of transit parameters. 

%Two questions naturally occur.
%
%(1) Are the limb darkening coefficients known well enough, that we can take them
%from some darkening tables and substitute them into Eqs. (1) and (4)? The 
%answer, as we will show in Section 3 is no.
%
%(2) What are the limitations and validity range of Eqs. (1) and (4) and of our
%knowledge on the limb darkening?

%To answer the second question, one must first point out two important limitations.

In Section 2 we study the differences between recent limb darkening tables.
In Section 3 we investigate the effect of the poor knowledge of the limb
darkening on the planet-to-stellar radius ratio. In Section 4 we check the
accuracy of the adjusted limb darkening coefficients and study this on
synthetic data. We call attention to the impact of stellar spots on the
transit fits and their importance in the observable limb darkening coefficients
in Section 5. We check the determinability of the limb darkening coefficients in
Section 6. Our conclusions can be found in Section 7.

\section{Comparison of theoretical predictions for limb darkening coefficients}

In Figure 1 the limb darkening coefficients predictions of Sing (2010) and
Claret \& Bloeman (2011) are plotted for three different stellar metallicities: $[M/H] =
-0.5, 0.0, 1.0$. It is obvious that the values given by these
different authors deviate from  each other by a value of $0.01-0.3$, depending
on which temperature region we consider. The usual range of the limb
darkening coefficients is between approximately $-0.1$ and $0.8$, so these 
 differences are considered as large.

We concentrate on the 3500-8000 K temperature range, because most of the
transiting exoplanets have been detected around main sequence stars with such 
 temperatures. Interestingly, we have better agreement between models in
the high-metallicity range ($[M/H]=1.0$). Here the coefficients agree with each
other, within $0.02$ at the low-temperature end, but over 7000 K, they differ from
each other by $0.05$.  Fortunately, this temperature region is less interesting
for transit hunters. At intermediate temperatures the predictions for limb
darkening coefficients do not differ by more than $0.01$ to $0.02$.

At lower metallicities the situation is completely different. Below $\sim
4500$K, the two lower metallicities are in complete disagreement, and the
relative differences between the predictions can be as high as $0.3$. Between
$\sim 5000$~K and $\sim 8000$~K the situation is still problematic, and in the case of
$[M/H] = 0.0$ the two tables generally differ by 0.2. Similar conclusions have also been reached by Claret \& Bloemen (2011). 

Some modellers fix the limb darkening coefficients according to these or other
tables when they fit a transit curve. As is clear from the aforementioned
discrepancies, in some temperature regions this is not a good strategy because
the values of the limb darkening coefficients are not consistent 
and it is not clear which are the correct ones.

When investigators adjust the limb darkening coefficients, then it is customary to
compare the results to some limb darkening coefficient table. Unfortunately,
most authors compare their results to only one table, and this may not be 
appropriate. It is necessary to compare the results to all available tables
because of the disagreements between tables.

Such a comparison was done by Southworth et al. (2007), among others. After modelling the
light curve of an eclipsing binary star, they compared the resulting limb
darkening coefficients to the tables of Diaz-Cordoves \& Claret (1995), Claret
(2000), and Claret \& Hauschildt (2003). The predicted values of these
three tables typically differ from each other by $\sim40-60$\% (see Table 5 of
Southworth 2007). In their paper the relative error of the resulting limb
darkening coefficients  was $\sim35$\%, and the limb darkening values observed 
were consistent with more than one theory. Consequently, they could not decide  
which theory describes the reality, so the uncertainty limits are
inconclusive for distinguishing the theories.

Claret (2009) concluds that no presently available limb darkening
calculation is able to reproduce the surface brightness distribution of HD 209
458. Claret (2008) also points out that the observations do not support
any of the limb darkening tables.

The differences between the different limb darkening tables are mainly due to
the adopted numerical methods and not to the atmosphere models themselves.
In addition, to have a clue to stellar atmosphere studies, it will be
necessary to compare the  differences between the observations and theory. This
gives another argument for why it is worthy and necessary to fit the limb darkening
coefficients.

%
%                                                One column figure
%----------------------------------------------------------- refl
   \begin{figure}
   \centering
   \includegraphics[width=6cm,angle=-90]{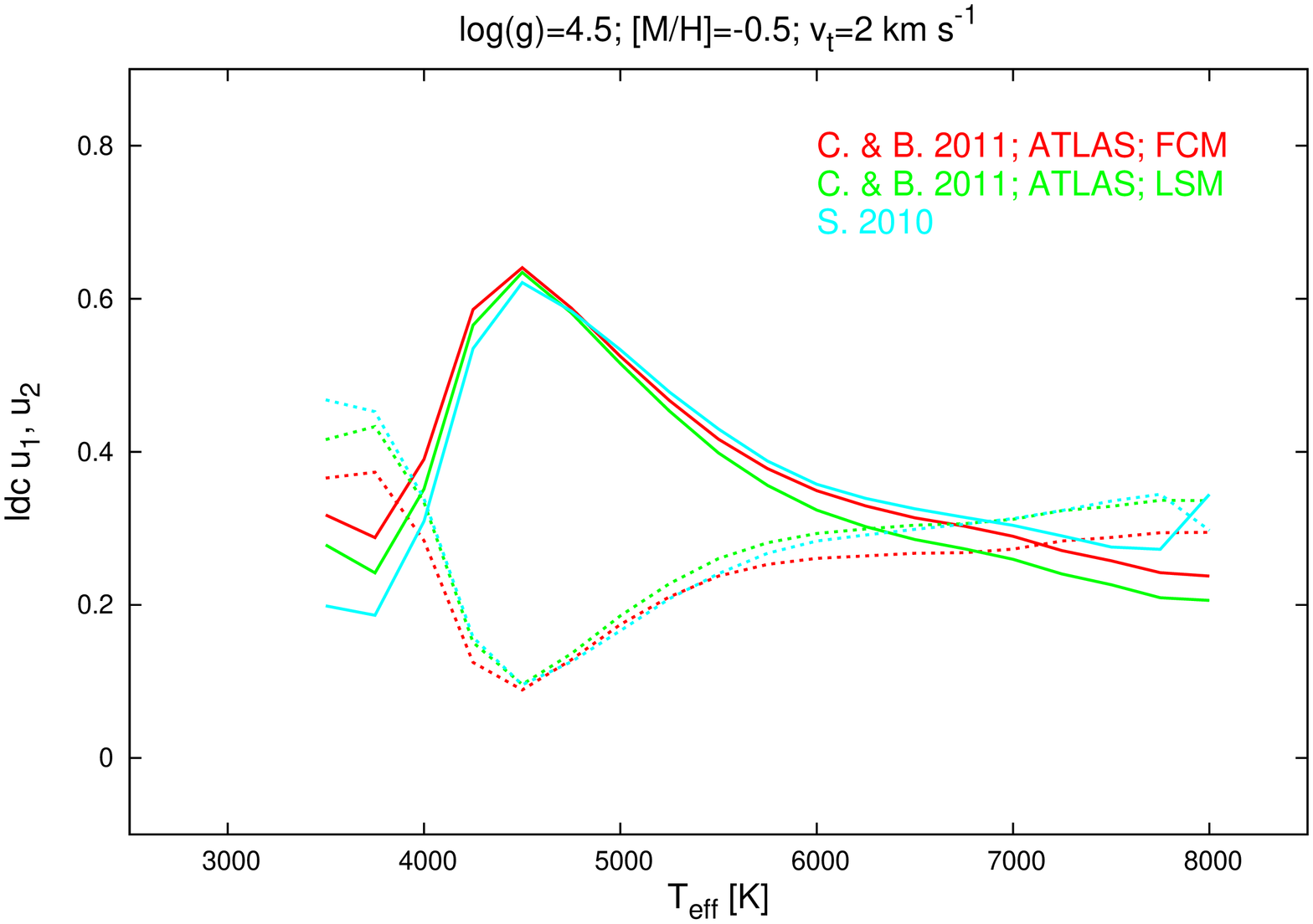}
   \includegraphics[width=6cm,angle=-90]{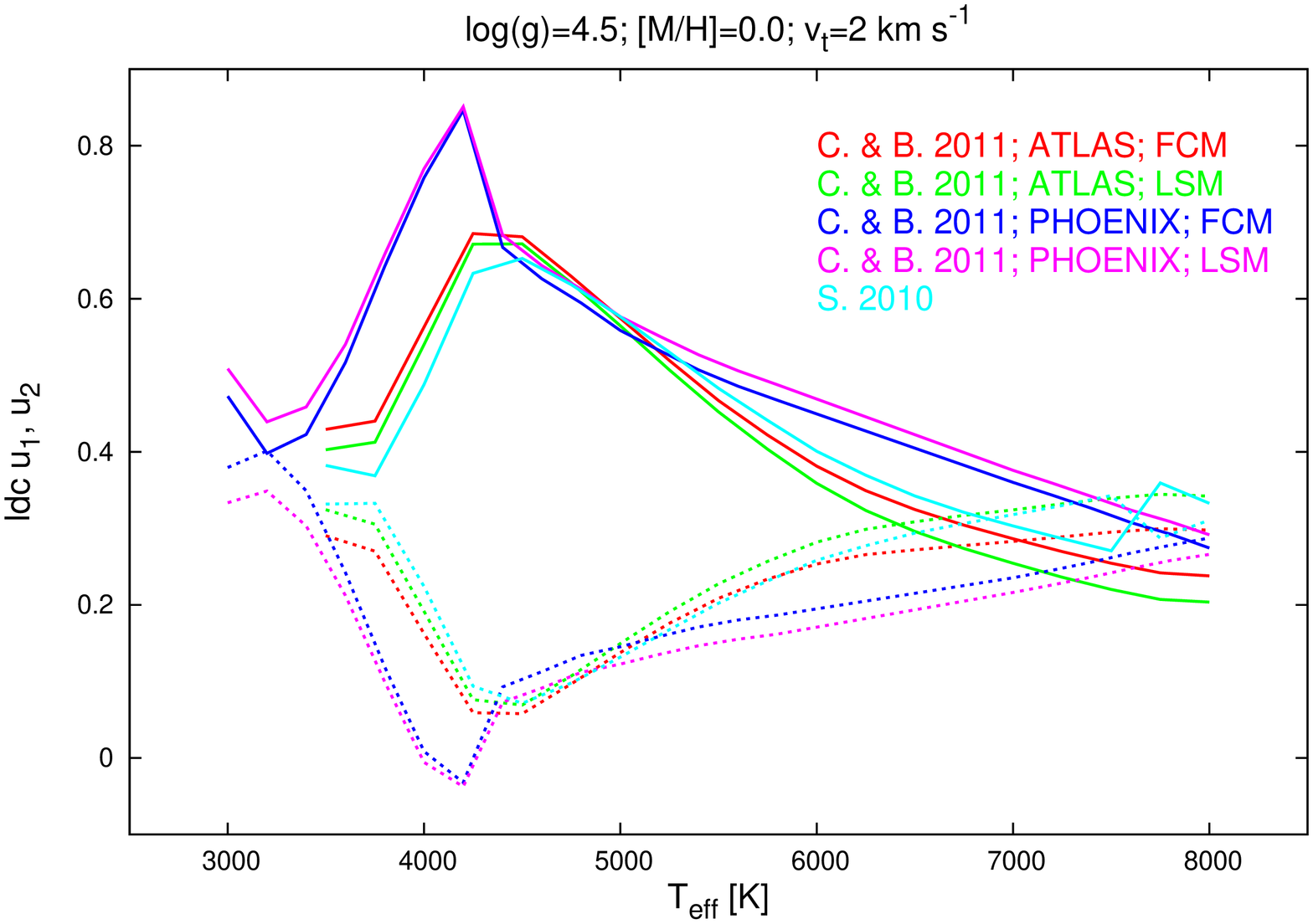}
   \includegraphics[width=6cm,angle=-90]{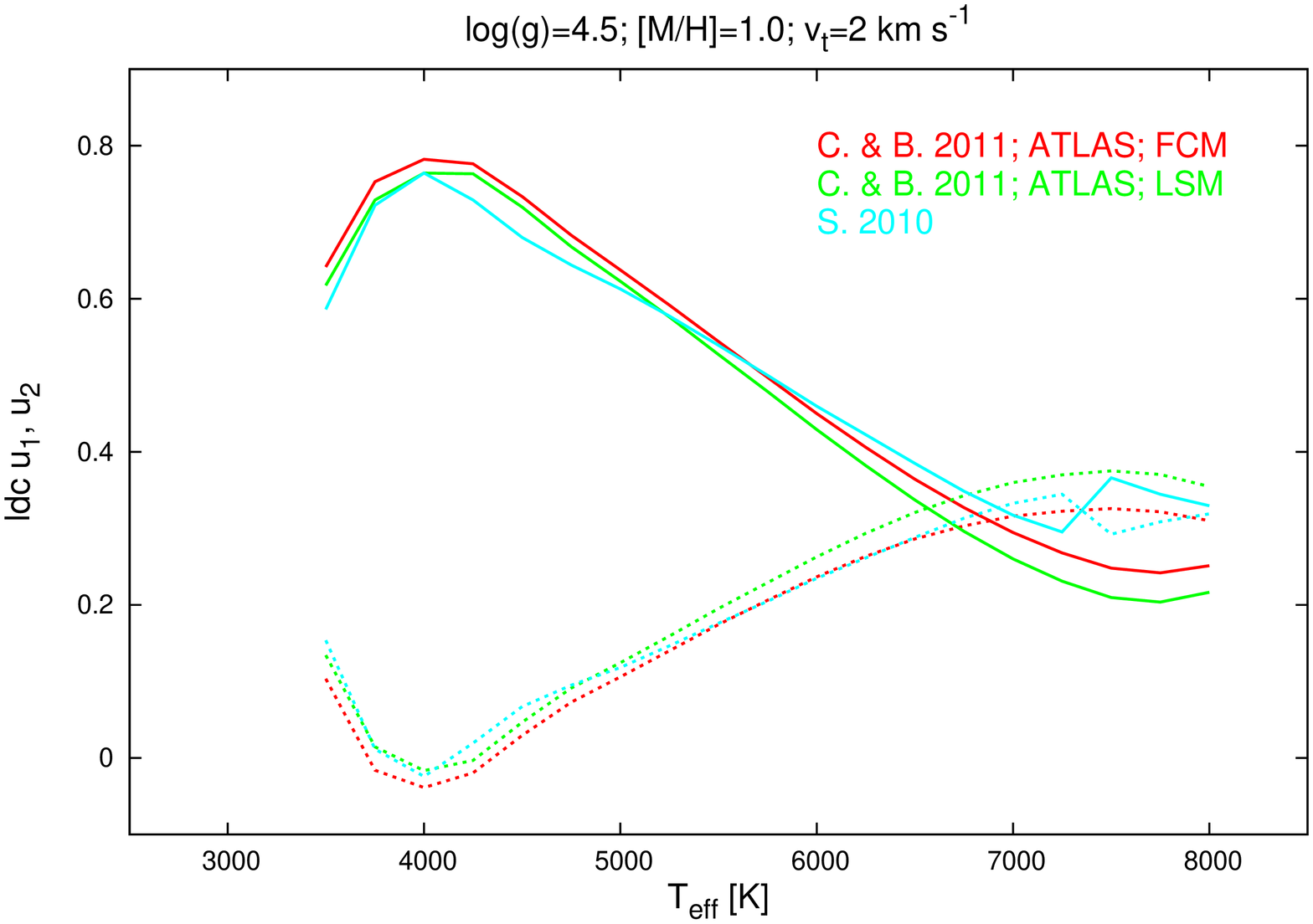}
      \caption{A comparison of 5 (3) different limb darkening models. We plotted only 
      the coefficients of the quadratic limb darkening-law.  From top
      to bottom one can see the models for different metallicities ($[M/H]=-0.5,
      ~0.0,~1.0$). For the sake of correct comparison, we used the same logg and
      turbulent velocity $v_t$ values for all the curves. Solid lines represent the
      $u_1$ coefficients, dotted lines the $u_2$ coefficients. The
      model of Sing (2010) is denoted by cyan lines. Red lines show the
      models of Claret \& Bloeman (2011), which are based on the ATLAS
      synthetic spectra and their `flux conservation method' (FCM). Green lines
      were obtained by them using the same spectra, but applying their `least
      square method' (LSM) for calculating the limb darkening
      coefficients. For solar metallicity ([M/H]=0.0), they also calculated the
      limb darkening coefficients with these two methods, but from the PHOENIX
      spectra, too (violet and magenta lines). The figure only hows the CoRoT white light passband
      coefficients; the discrepancies are similar for Kepler's response
      function and for other photometric systems.}
         \label{reflexion}
   \end{figure}
%
%______________________________________________________________
%

\section{The effect of poor knowledge of fixed limb darkening coefficients on the radius
determination}

Here we investigate the hidden inaccuracy caused by fixing the limb darkening
coefficients during the transit-fitting procedure. Our goal is to calculate the
error in $k$ caused by the theoretical uncertainties of the limb darkening
coefficients. We assume the time being that the normalized transit depth $\Delta
F/F$ is perfectly known since we are only interested in the error caused by the
inconsistent limb darkening tables.

Let $\Delta u_1$ and $\Delta u_2$ be the uncertainties of the 
theoretically calculated linear and the quadratic limb darkening 
coefficients, respectively. These uncertainties have four sources: 
(i) the theories have some uncertainties. These kinds of uncertainties 
are not well known, but as a first approximation, we can assume that 
their order of magnitude is equal to the differences between different 
theories as mentioned in Section 2. (ii) The input stellar parameters 
($T_{eff}$, $log g$, metallicity) have uncertainties that will be 
reflected in the prediction of limb darkening coefficients. The 
uncertainties of stellar parameters usually cause small 
uncertainties in the prediction of the limb darkening coefficients, and  
for instance, Deleuil et al. (2012) found that the uncertainty of the 
predicted limb darkening coefficients was $\pm0.0168$ when they 
propagated the stellar parameter uncertainties to the limb darkening 
prediction. In most of the cases the stellar paramater uncertainties 
cause less than $\pm0.03$ uncertainty in the prediction of the {limb 
darkening} coefficients. (iii) Although the spectroscopic effective 
temperature determination is not significantly affected by stellar spots 
in the case of a solar-like activity, the stellar surface effective 
temperature determination can be systematically affected when the spot 
coverage is about 10-20\% (A. Hatzes, priv. comm., Ribas et al. 2008), 
and this can lead to systematic errors in the limb darkening predictions 
and to underestimation of the uncertainties of such a prediction. Thus, the 
uncertainties mentioned in point (ii) might be underestimated. It is quite 
difficult to characterize these kinds of error sources, but fortunately 
they have an impact only for very active stars, which are usually avoided 
by planet hunters. (A remarkable exception is CoRoT-2.) (iv) The 
inhomogeneties of the stellar surface temperature distribution (e.g. 
spots, faculae, gravity darkening-effects, etc.) are not included in  
the prediction of limb darkening coefficients.

Now we consider the uncertainty in the planet-to-stellar radius ratio caused by
the uncertainty of the limb darkening coefficients. After straightforward
calculation we have (cf. Eq. (B1)):
\begin{equation}
\frac{\Delta k}{k} = \frac{1}{2} \left(\frac{\Delta u_1 \delta + \Delta u_2 \delta^2}{1-u_1 \delta - u_2 \delta^2 } +\frac{2\Delta u_1 + \Delta u_2 }{6-2u_1 - u_2} \right)
\end{equation}
where we abbreviated $\delta = 1 - \mu$, and $\Delta x$ means the uncertainty
of quantity $x$. The variation in the relative error in the radius ratio of the star and the
planet from Eq. (1) is shown in Figure 2.

%
%                                                One column figure
%----------------------------------------------------------- refl
   \begin{figure*}
   \centering
   \includegraphics[width=8cm]{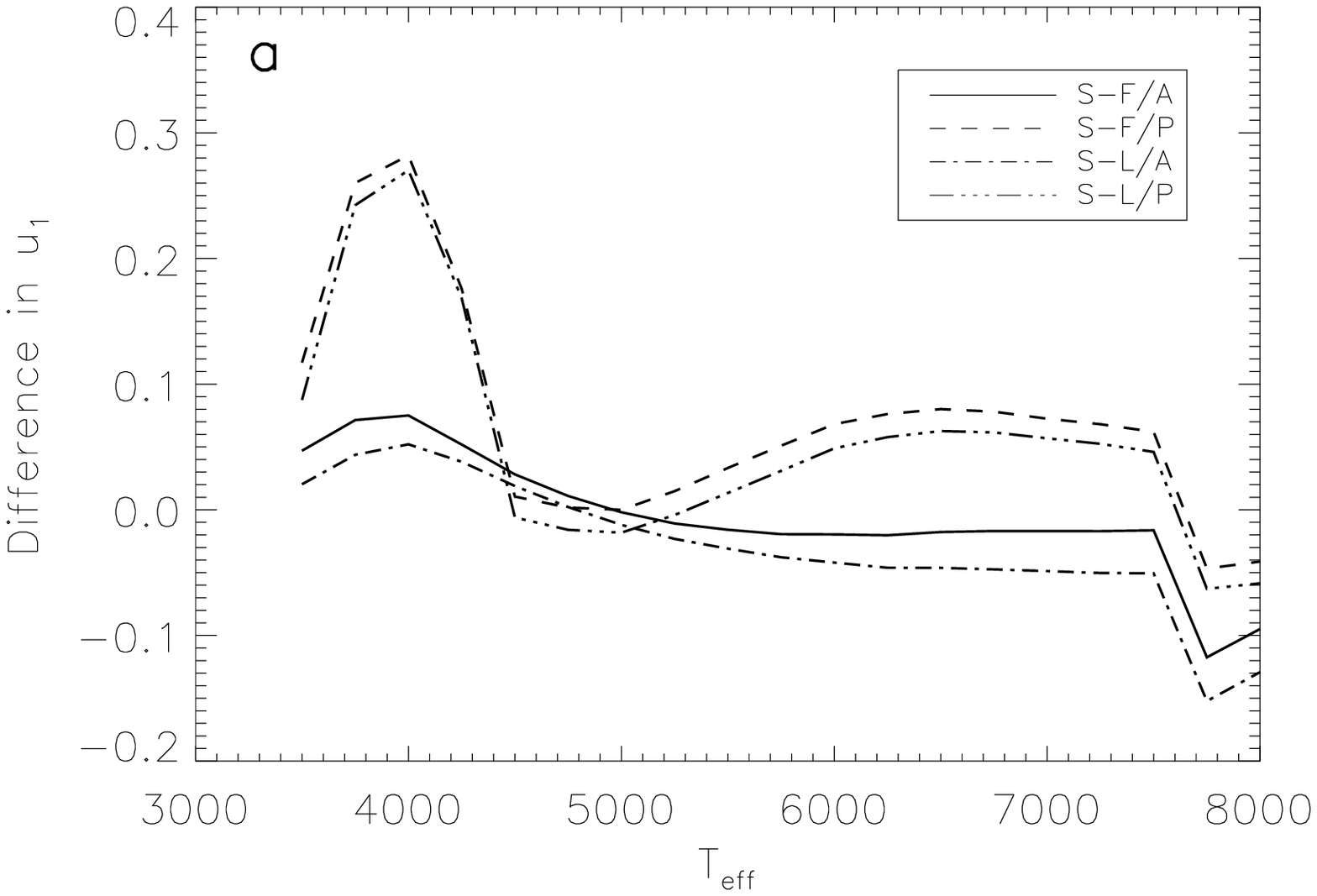}
   \includegraphics[width=8cm]{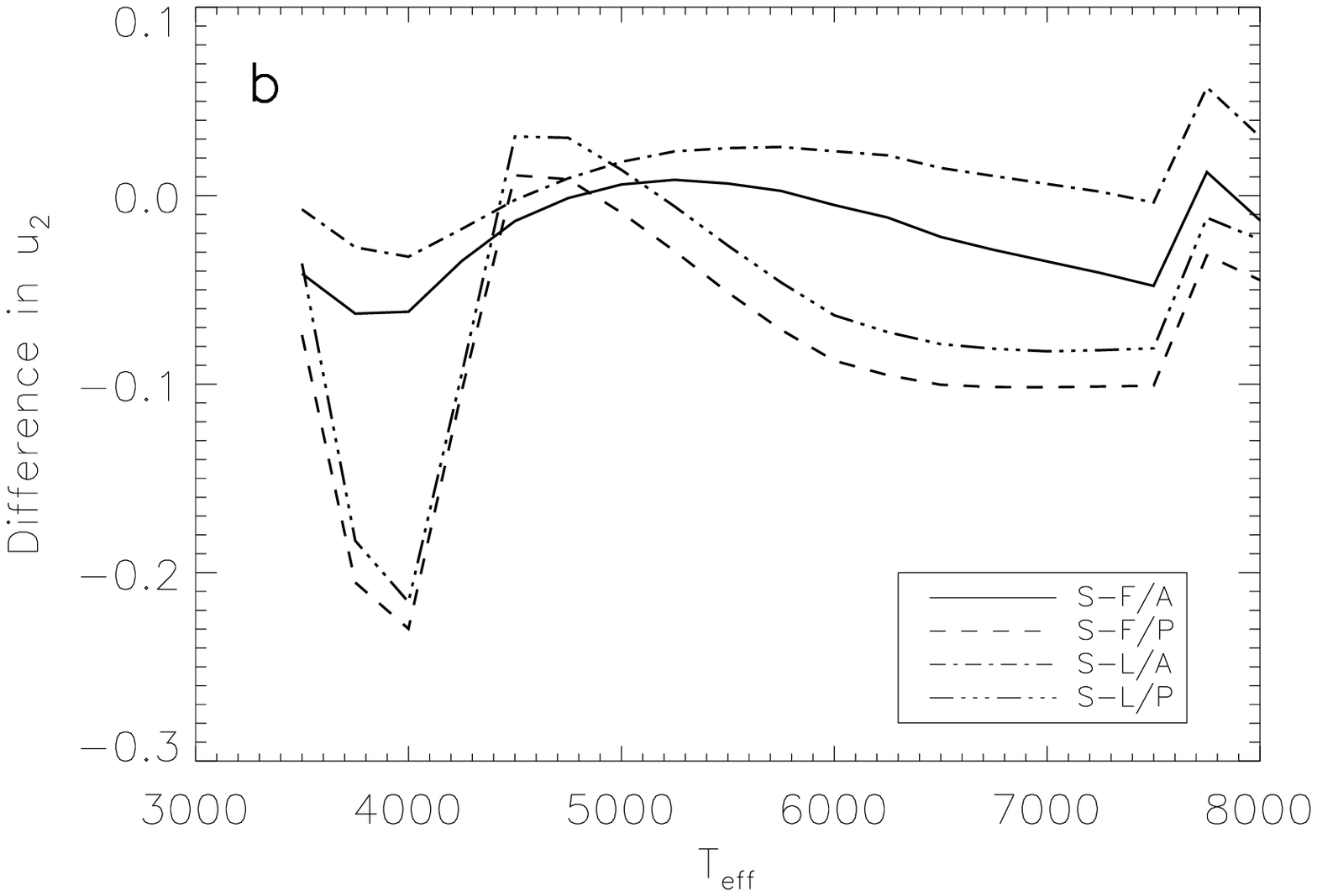}
   \includegraphics[width=8cm]{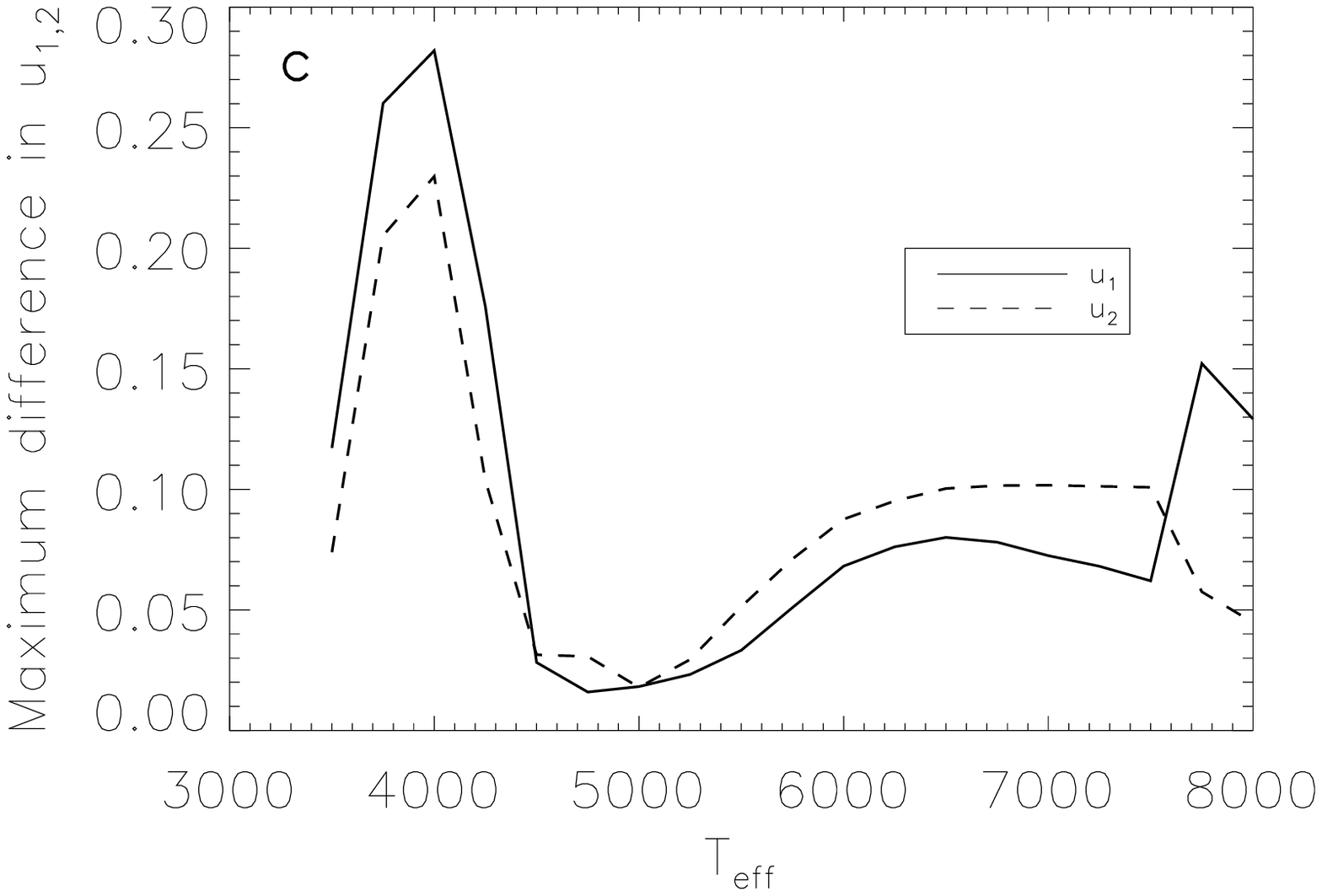}
   \includegraphics[width=8cm]{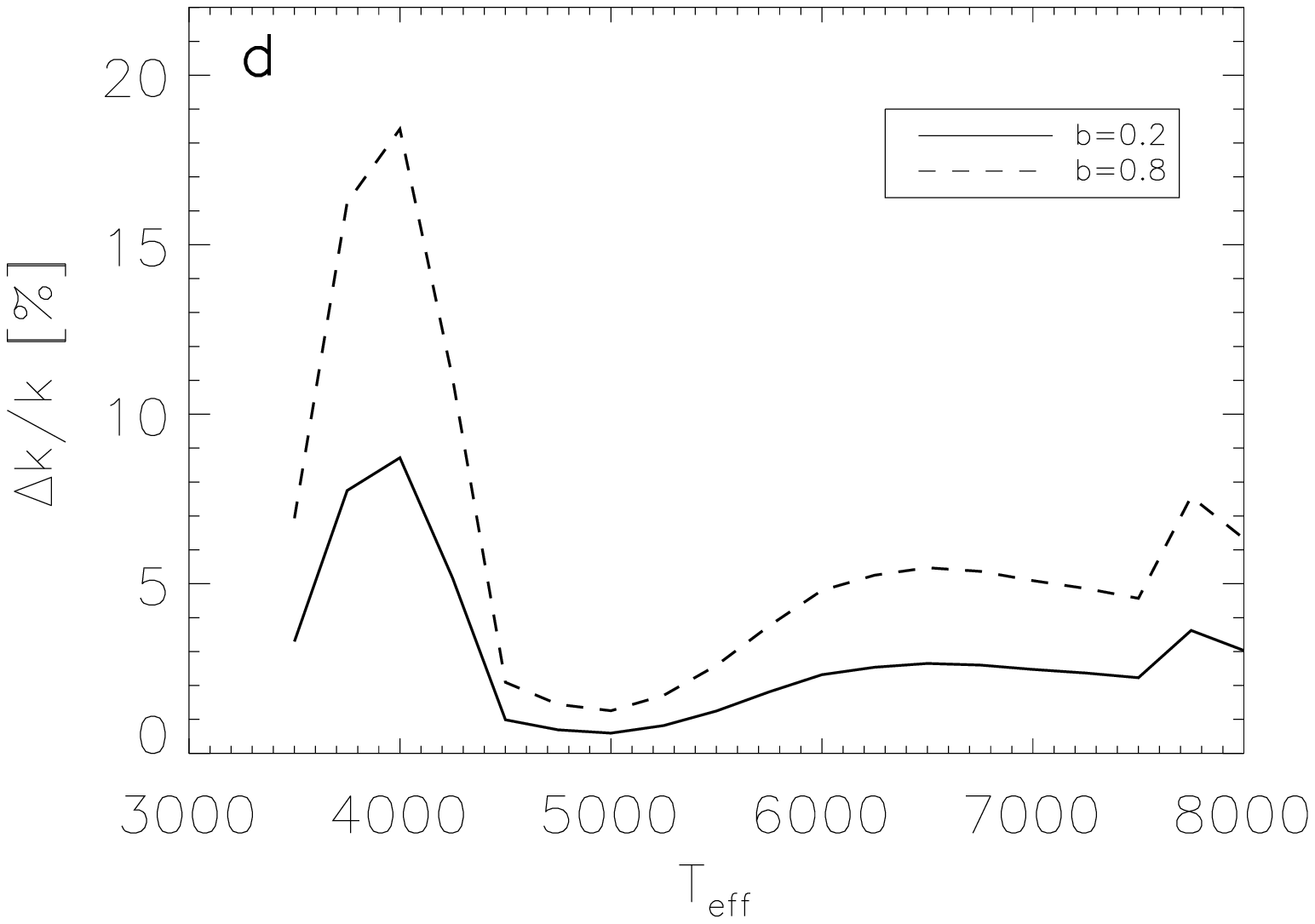}
      \caption{a): The difference of the theoretically predicted limb
      darkening coefficient $u_1$ as a function of the stellar effective surface
      temperature between Sing (2010) (S) and different models of Claret \& 
      Bloemen (2011). F means the flux conservation method, L means
      the least square method of Claret \& Bloemen (2011). A and P denote the
      ATLAS and the PHOENIX models that they used for creating of their 
      limb darkening tables. b): the same for the $u_2$ quadratic coefficients. c):
      the absolute value of the maximum differences at a given temperature
      between Sing (2010) and some of the Claret \& Bloemen (2011) models. The
      curves in this panel are upper limits, because we calculated the
      differences between the different limb darkening tables in a conservative
      way; i.e., we searched for the biggest difference between Sing's table and
      any of the Claret \& Bloemen's table at the given temperature. 
      d): visualization of the meaning of Eq. (1): relative error of the
      radius ratio of the star and the planet $k=R_{planet} / R_{star}$ vs
      effective surface temperature of the stellar surface when the limb
      darkening coefficients are fixed. The $\Delta u_1$, $\Delta u_2$
      uncertainties are taken from panel c. The two curves are valid for a rather
      central ($b=0.2$) and for a rather grazing ($b=0.8$) transit. For
      these plots we used the CoRoT passband limb darkening coefficients of 
      Sing (2010) and Claret \& Bloemen (2011), and we assumed $log g=4.50$, solar
      metallicity and a turbulent velocity of 2km/s. However, the main result
      does not change significantly for other input
      parameters or passbands. The uncertainty of the parameters were estimated as the
      difference between the two limb darkening tables. In certain temperature
      regions we are far from the required precision in the radius ratio 
      determination.}
         \label{k-uncertainty}
   \end{figure*}
%
%______________________________________________________________
%

First we consider the temperature region $5000 \mathrm{K} < T < 7500$~K. Here
the different limb darkening theories agree more closely. Substituting the
uncertainties of the theory discussed in the previous section, we can easily
conclude that the uncertainties in the relative radius ratio of the planet and
the star varies between 1\% and 10\%, but it can be around 8\% at the high temperature. 
Below 5000 K, where the curves of different theories are very divergent, we have
uncertainties of up to 20\% in the radius ratio.
%\footnote{For this estimation we used du1=du2=0.3, u1=0.6, u2=0.2, and 
%delta=1-mu, mu=sqrt{1-b^2} and b varied between 0 amd 1}.
These uncertainties only come from the inconsistencies of the theoretical
limb darkening calculations. The aforementioned two additional factors increase
 these uncertainties further, but one needs a case-to-case study to characterize 
them.

These uncertainties are unacceptable if radii and therefore internal structure
of the planets are being studied, because to distinguish between different planet
models and to study them in detail we require $\pm1$\% precision in planetary
radii below five Earth-masses. The consistency of the limb darkening-tables should be improved
in all temperature regions and should be checked via observations, to obtain
 more concordant limb darkening tables in the near future.

Since we do not have reliable limb darkening theories it seems questionable to
fix the limb darkening coefficients for transit light curve analysis. For
instance, CoRoT and Kepler are able to determine the transit depth with a
precision of $10^{-4}$ in general or sometimes even better, which can be
translated to 0.5\% relative error for the radius ratio $k$ -- in the absence of
limb darkening. Since the error due to the poor knowledge of the limb darkening
may be as big as 20\% at low host-star surface temperatures ($T<5000$K), this
can cause $\sim40-50$ times larger error in the radius than the error stemming
from the quality of photometry alone! This error source is also two to ten times
bigger at the higher host-star's surface temperatures ($T>5000K$) than the error
stemming from the photometry. This is the case when one fixes the
limb darkening coefficients during the transit light curve fit procedure.

We add that these numbers come from the optimistic case;  
i.e., uncertainty in the stellar radius can be neglected. But this is not 
the case. The aforementioned errors are the errors stemming from the 
planet-to-stellar radius ratio determination alone. When this radius 
ratio is transformed to absolute dimensions of the planet via $R_{planet} 
= k R_{star}$, then the error bars should be increased more accordingly, by the uncertainties in the stellar radius. The 
stellar radius is usually obtained by comparing stellar evolutionary 
models to the spectroscopically measured quantities, such as $T_{eff}$ and 
$\log g$ of the star. It seems that the most careful spectroscopic 
studies of the host stars are able to yield the stellar radius with a 
precision not better than $\sim 5$\% (e.g. Bruntt et al. 2010; see 
also the discussion in Torres et al. 2012). However, the final error 
budget also depends on the uncertainties and systematics of the stellar 
evolutionary models used. For instance, the radii of stars below 1 
$M_\odot$, predicted by current evolutionary models, are not supported 
by observations (see e.g. Clausen et al. 2009 and references therein). 
It is beyond the scope of this paper to characterize these kinds of 
error sources, we limit our study only to the effect of limb darkening 
on the planet-to-stellar radius ratio, and therefore we refer to Torres 
et al. (2010), where the uncertainties of the used stellar models and the 
derived stellar parameters are discussed. We also mention that future 
projects, like PLATO, are expected to provide the stellar masses and 
radii with less than couple of percent relative error via 
asteroseismological studies. Details of this latter method and its 
uncertainties are discussed e.g. in Catala (2009).

\section{Adjusting the limb darkening coefficients}

Here we study how precisely our fitting procedure is able to determine the different
transit parameters. For this purpose we created 2000 synthetic light
curves with the subroutines of Mandel \& Agol (2002). We assumed a circular
orbit for all of these curves. The number of the simulated photometric data
points inside the transit were selected randomly between 500 and 5000. Then we
added Gaussian-type random noise to these light curves with zero mean and
$1\sigma$ scatter, and we chose randomly $\sigma$ such that our final light
curves have S/N between 1 and 1000. The S/N is
defined as follows: $S/N = k^2 / \sigma $. We did not study the effect of red
noise, systematic errors, etc., since our aim is to find what minimum
accuracy is needed for precise parameter determination. The input parameters of
these synthetic light curves were randomly chosen between the following limits:
$b=0...1$, $k=0...0.3$, $a/R_s = 5...50$, $u_+ = 0...1$ and $u_- = -0.1...1.0$,
respectively. Here, $a$ is the semi-major axis, $R_s$ the stellar radius - $a/R_s$
are often called scaled semi-major axes -, $b$ is the impact parameter, and $u_+
= u_1 + u_2$ and $u_- = u_1 - u_2$. The use of these combinations of the limb darkening coefficients was suggested by Brown et al. (2001) and they aim to avoid 
degeneracies between $u_1$ and $u_2$. We selected $u_+$ and $u_-$
randomly between the aforementioned limits. This means that they are not related
in this test to any stellar properties or theoretical considerations, and thus
there is no link between them; i.e., they are independent of each other in this
test. %This becomes important in Section 7 where we show that we can determine
%the limb darkening coefficients even if they disagree with the theoretical
%predictions.

In the next step we modelled these artificial light curves with our fitting
procedure, which is based on a genetic algorithm (Geem et al. 2011). The free
parameters were the impact parameter $b$, $k$, $a/Rs$ ratio and $u_+$. We fixed
the value of $u_-$ at 13 different values at $-0.3, -0.2, -0.1, ..., 0.9, 1.0$,
and then we selected the best solution (according to the $\chi^2$-values). Then, 
this best solution was refined by allowing $u_-$ to vary in a narrower range ($\pm$0.1). After that we plotted the standard devations of the 
differences between the value obtained from the light curve solution and the
original input value, normalized to the input value vs the S/N ratio as it is
plotted in Figure 3:
\begin{equation}
s = \mathrm{STD~DEV}\left(\frac{|\mathrm{modelled~value}-\mathrm{input~ value}|}{\mathrm{input~ value}}\right). \nonumber
\end{equation}
This $s$-value has been plotted in the different panels of Figure 3 against the
S/N of the simulated light curves. We draw conclusions from this figure as
follows. We need at least S/N=25 and 6 to determine the limb darkening
coefficients and the scaled semi-major axis with $\pm 5$\% uncertainty. To
determine $k$ with $1$\% accuracy, we need at least S/N=$\sim$50. To determine
$b$ with $10$\% relative accuracy, we need S/N$=\sim100$ at least. We also 
found that when $b>0.85$ the solution becomes unstable and far from the input
values, so we removed them from this analysis.

%However,
%since all exoplanets found sofar have an inclination close to $90^\circ$, we do %not need
%much higher accuracy in the impact parameters (i.e. the inclinations) to
%determine the planet masses better than now. (Note that $b$ is related to the %inclination $i$ via $b = a \cos i (1-e^2) / R_s (1+e\sin \omega)$,where $a$, %$e$, $\omega$ are the semi-major axis, eccentricity and the argument of %periastron, respectively, and radial velocity gives only the $M_p \sin i$ %parameter combination for the planetary mass $M_p$.) 

%
%                                                Two column figure
%----------------------------------------------------------- refl
   \begin{figure}
   \centering
   \includegraphics[width=6.4cm]{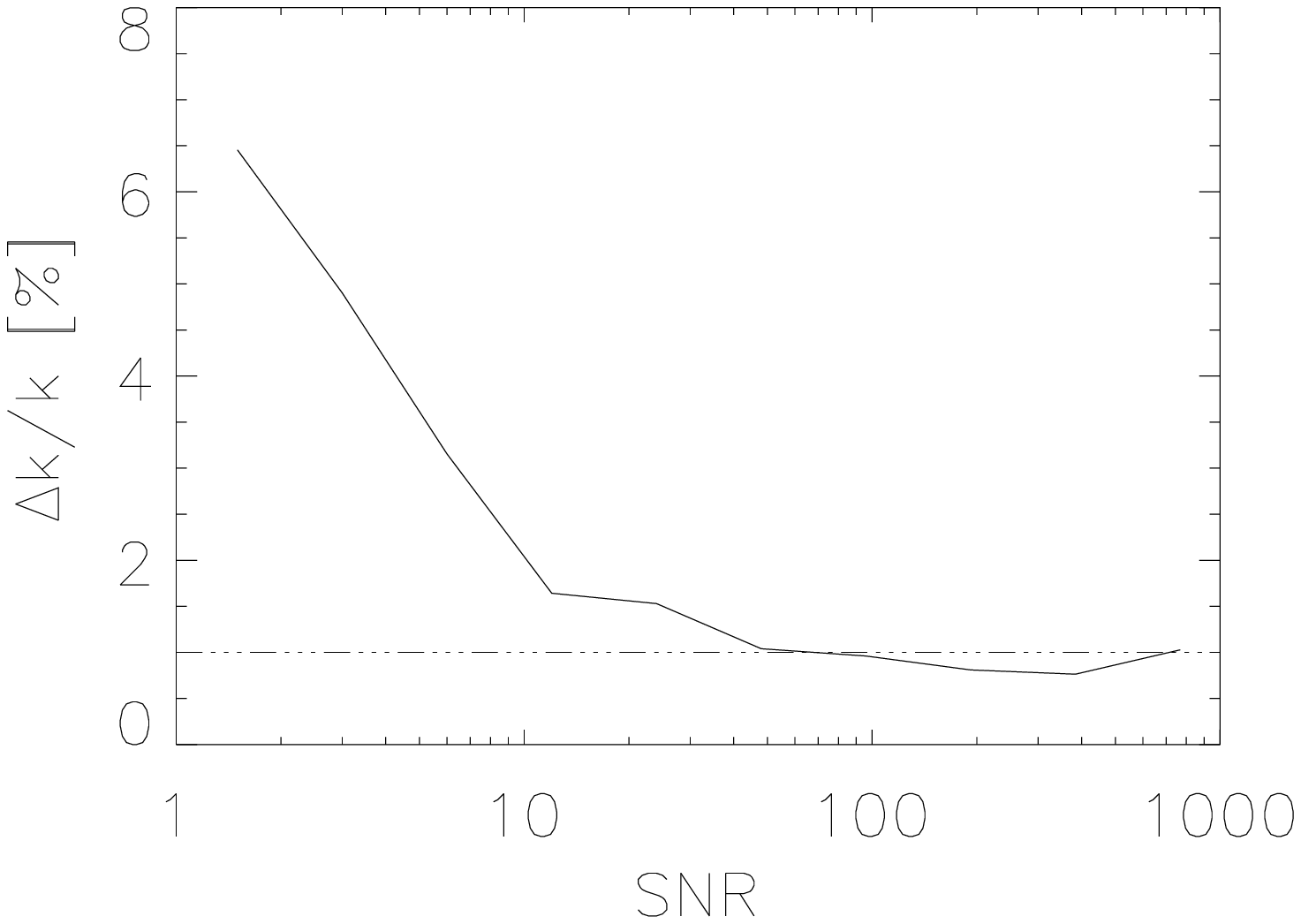}
   \includegraphics[width=6.4cm]{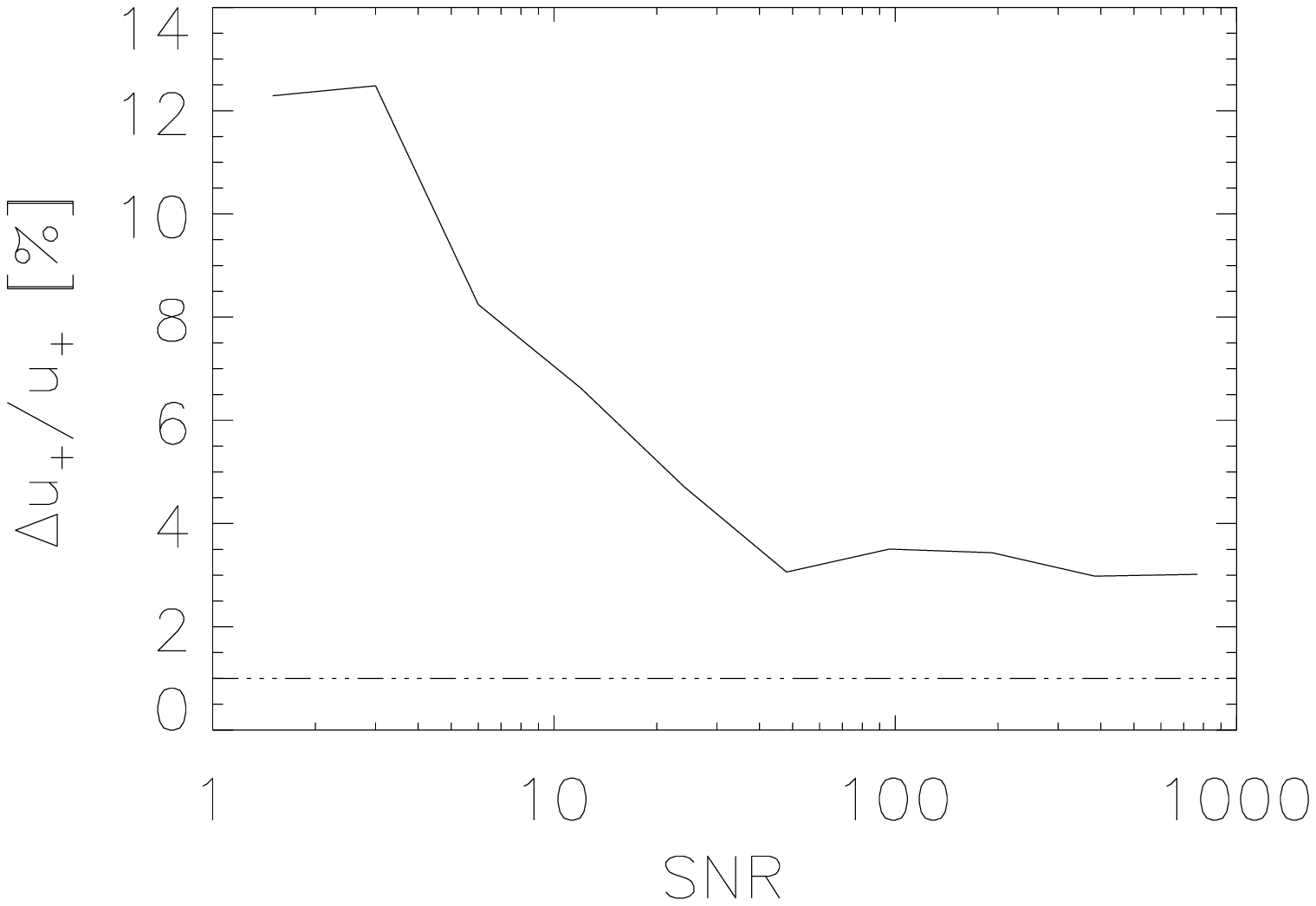}
   \includegraphics[width=6.4cm]{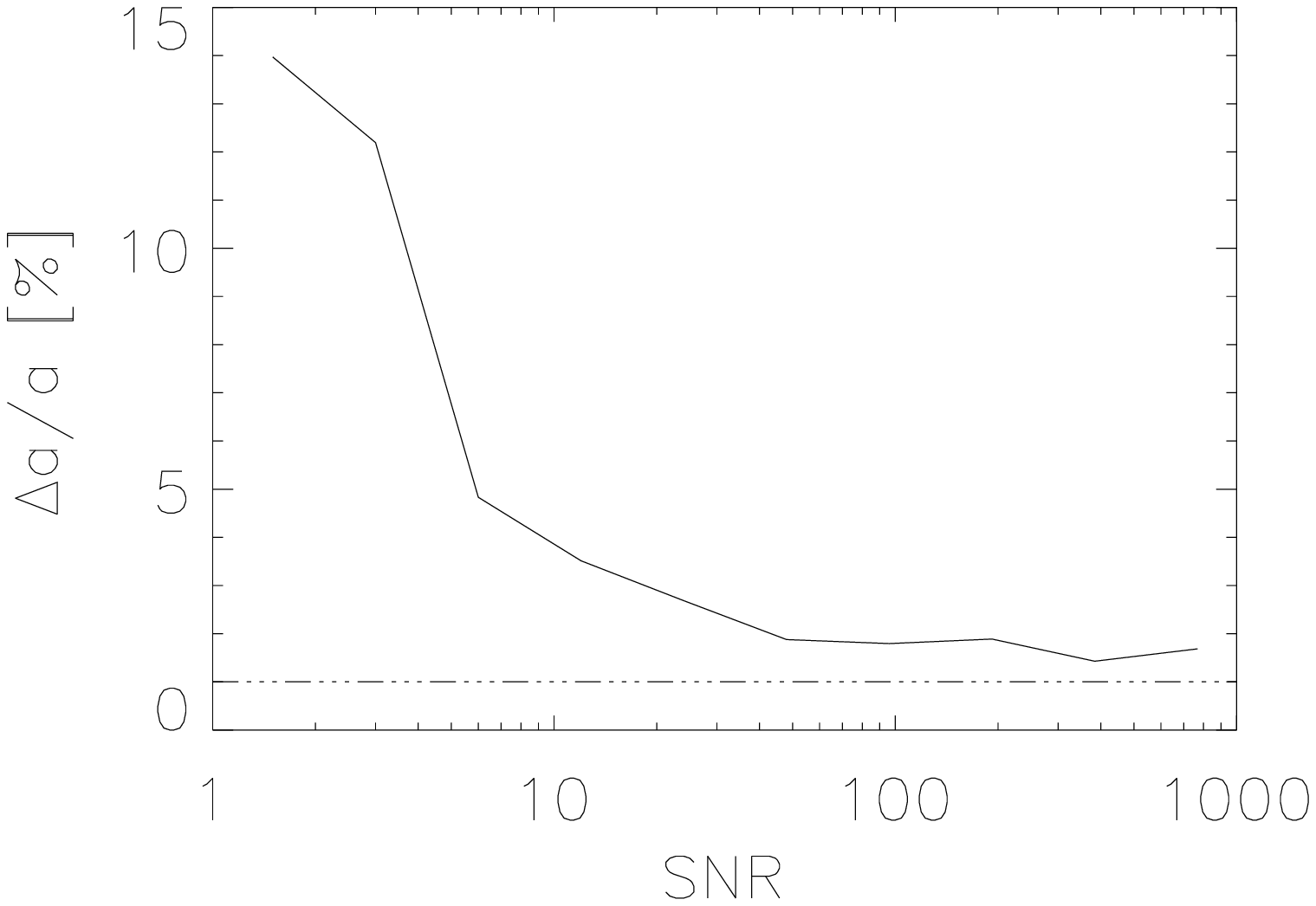}
   \includegraphics[width=6.4cm]{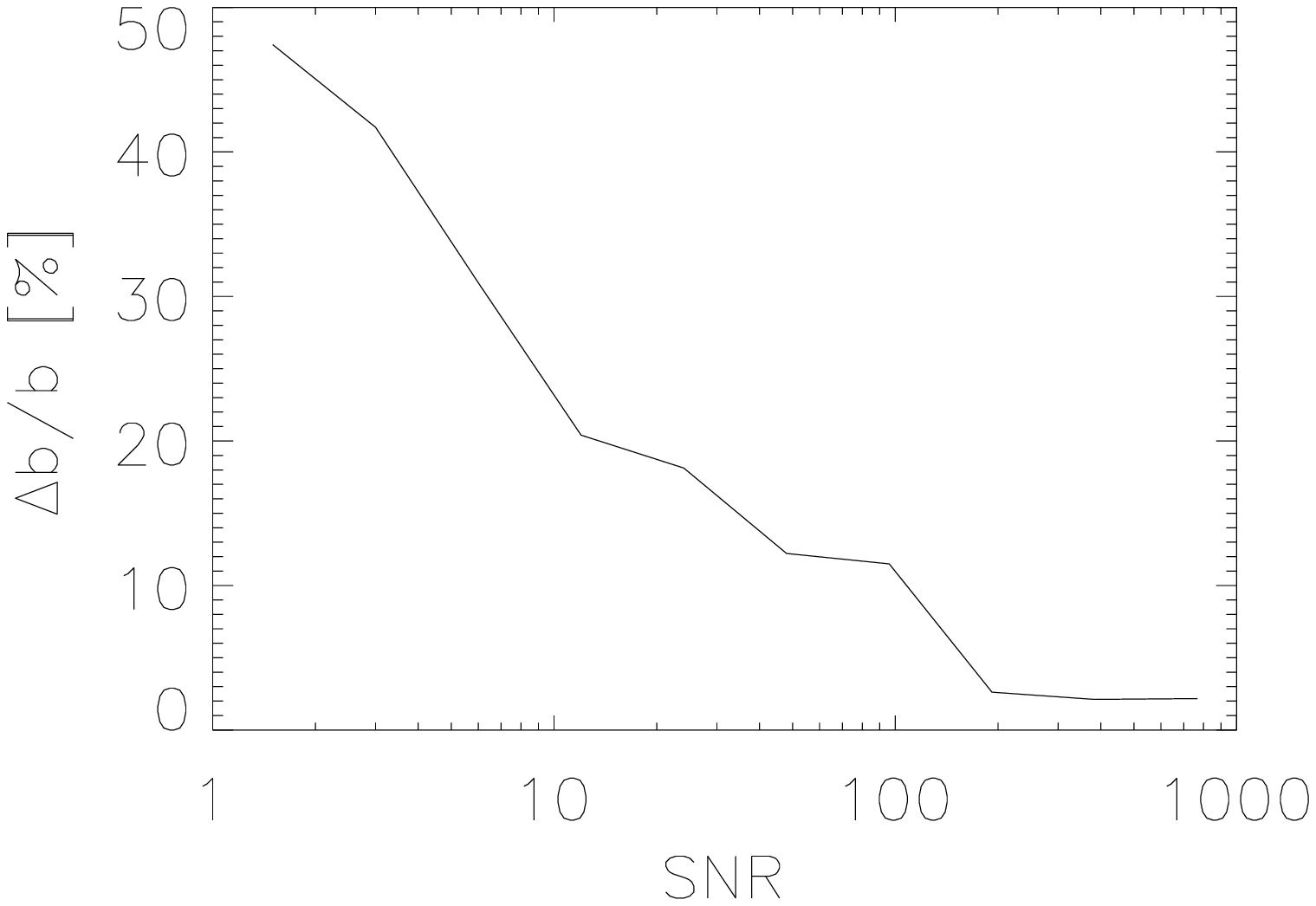}
      \caption{Results of the light curve modelling test. On the x-axis the S/N
      quantity is defined in the text. y-axes are the relative uncertainties in
      radius ratio $k$, $u_+$ combination of the two quadratic limb
      darkening coefficients used, in the scaled semi-major axis (for brevity we used
      the notation of $a$ only instead of $a/R_s$ in the third panel from top)
      and in the impact parameter $b$, from top to bottom, expressed in
      percentages. The solid lines show the relative uncertainties defined
      as the standard deviations of the differences of the modelled and the 
      input values normalized to the input values at certain S/N-values.}
         \label{Test}
   \end{figure}
%
%______________________________________________________________

\section{Effect of stellar spots on the limb darkening}

The result of a fit with adjusted limb darkening coefficients can lead to
unreasonable limb darkening values, e.g. Southworth (2011) accepted a worse
solution for CoRoT-13b in the sense of the quality of the fit (measured by the
$\chi^2$-values), because when he adjusted both limb  darkening coefficients,
he obtained `unphysical' values, i.e. far from the theoretically predicted
range, sometimes causing `limb brightening' instead of limb darkening or
negative fluxes at the edge. What can we say about such cases?

We would like to point out that we employed an indirect assumption in the previous
sections: the effective temperature of the stellar surface is assumed to be
the same at each point of the surface and therefore the stellar surface brightness
distribution is only modified by limb darkening. The same assumption is used
when the theoretical limb darkening coefficients are calculated, and almost every transit light
curve investigator uses the same assumption. However, there are several factors
that modify the {\it local} effective temperature on the stellar surface. Such
effects are e.g. stellar rotation (i.e. gravity darkening, see e.g. von Zeipel
1924; Twigg \& Raffert 1980; Barnes 2009, Espinoza \& Rieutord 2011; Claret 2012), reflection effect (although we
can neglect this for the star in a star-planet system), and stellar activity phenomena: spots, plage-area, faculae and
flares. Some of these are transient phenomena, e.g. the flares, but others have
similar or much longer timescales than the length of the transits (e.g. the
spots and faculae). They can be regarded as an additional, but constant surface
brightness distortion during a transit. Constant here means that we assume that the number, temperature, size, astrographic
longitude, and latitude of the distortion(s) do not change around and during a
transit.  %However, to present our conception the used assumption on the
The constancy of the aforementioned quantities are reasonable.

These distortions mean that the local effective surface temperature where they occur is lower for stellar spots and higher for faculae
than that of the undistorted stellar surface. As is well known, the
limb darkening coefficient is different for different surface temperatures. {\it
This means that the apparent stellar disc cannot be characterized by only one
effective surface temperature value, and that is why the surface brightness
distribution cannot be characterized by only using the limb darkening
coefficients due to only one $T_\mathrm{eff}$.}

Djurasevic (1992) already took into account that the undistorted and the
distorted areas have different limb darkening coefficients, when he developed a
code to analyse eclipsing binary star light curves. The corresponding limb
darkening coefficients were taken from theoretical approximations available at
that time. We would need a similar approach for the transit modelling. However,
we suggest avoiding exactly the same approach for the following reasons.

\begin{itemize}

\item[a)] Different tables give different limb darkening coefficients, which 
is why the results can be different, and maybe the derived spot parameters will
depend on the assumed limb darkening coefficients;

\item[b)] The spot distribution is not well known, and the spot
solution is often highly degenerated so the reliability of the spot solution can be criticized if only
one-colour photometry is available. Consequently it is hard to include the spots in a
modelling tool.

\end{itemize}

\subsection{Effect of a single active area}

%In Section 6 we established the signal-to-noise ratio limits for the
%determinability of different transit parameters, and we have also shown that
%even if the limb darkening coefficients cannot be determined from the lower
%signal-to-noise transit light curves, the other important transit parameters,
%$k$, $b$ and $a/R_{star}$ can still be obtained with satisfactory accuracy.
%However, we also pointed out above that the presence of stellar spots and
%faculae can disturb the fit. Hereafter we show that in this case the parameters
%$b$, $k$ and $a/R_s$ remains still well-determinable, their precision is
%satisfactory ($<1$\%) if the signal-to-noise ratio is high enough, only the
%value of the limb darkening coefficients will change and will agree no longer
%with the theoretical values even for high SNR light curves.

To investigate the effect of stellar spots and faculae (hereafter we
call both spots, including spots that can have higher temperatures than the
surface, i.e. ``bright spots'', describing the characteristics of
faculae that they have a higher temperature than the normal photospheric area), we classify the spots as follows. Spots that are not eclipsed by the
planet we define as Type I spots and spots that are eclipsed by the planet as Type II
spots.

The effect of Type II spots can be taken into account because the
size and location of the spots can be determined with high accuracy, and that is
why removing of their contribution from the light curve causes no problem, at
least in principle (Silva-Valio \& Lanza 2011; Sanchis-Ojeda \& Winn 2011;
D\'esert et al. 2011).

Type I spots are more complicated. Those that cause light 
curve modulations can be removed safely with baseline fits and
baseline corrections using the out-of-transit points in the small vicinity of
transits. But there are also spots that cause no light curve modulations, e.g. a polar
spot. In addition, Jackson \& Jeffries (2012) propose, based
on spectral evidence, that a considerable  amount of stars exhibit numerous (up
to $\sim 5000$), small (not bigger than 2-3 astrographic degrees) dark spots that are 
axysymmetrically distributed on the stellar surface, reaching $\sim50$\%
spot-coverage, which led to no observable light curve modulation over the
current detection limit in spite of their high activity level. If this is true
then we can expect that even photometrically quiet stars can produce strange
limb darkening coeffients as we will show.

%
%                                                One column figure
%----------------------------------------------------------- refl
   \begin{figure}
   \centering
   \includegraphics[width=8cm]{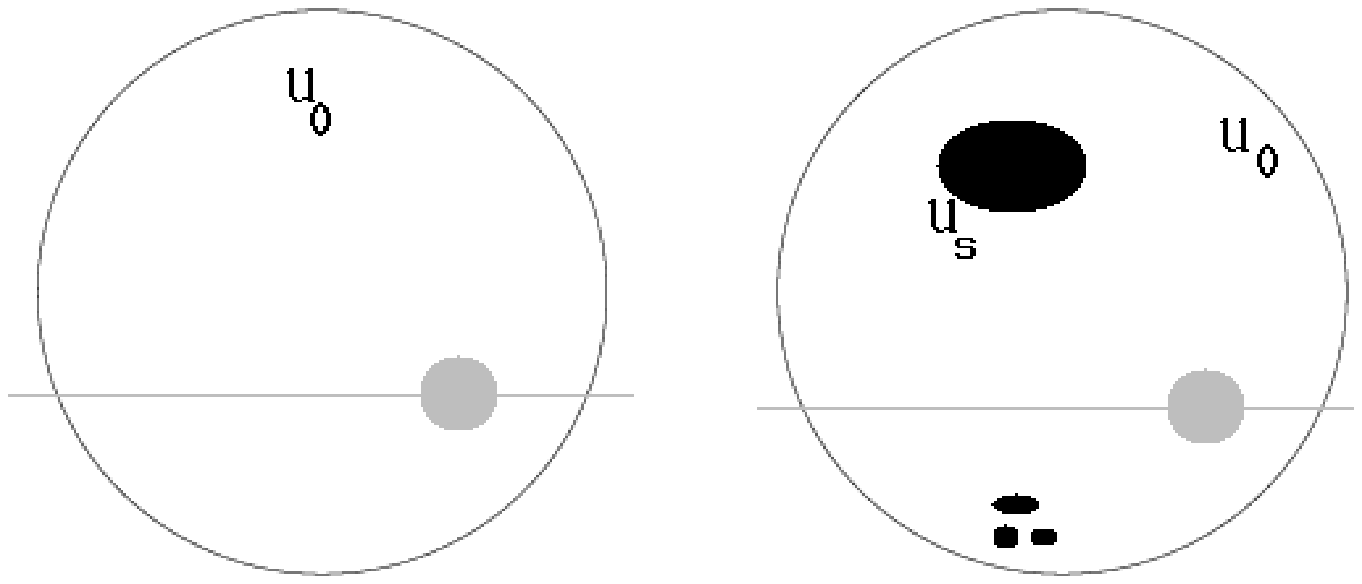}
      \caption{Illustration of the effect of Type I spots. Left: the planet crosses
      an unmaculated star that is characterized with some limb darkening
      coefficient $u_0$. Right: the planet crosses the appareant stellar disc of a
      spotted star, where the spots and the planet have different impact parameters, 
      as well as the stellar photosphere and the spots have different limb
      darkening coefficients ($u_0$, $u_s$). Grey area is the planet, black ellipses represent the spots.}
         \label{reflexion}
   \end{figure}
%
%______________________________________________________________
%

Figure 4 illustrates the situation. When the planet moves in front of an
unmaculated star, the observable maximum flux change will be
\begin{equation}
\frac{\Delta F}{F_0} = k^2 \frac{1-u_{10}(1-\mu_p)-u_{20}(1-\mu_p)^2}{1-\frac{u_{10}}{3} - \frac{u_{20}}{6}} 
\end{equation}
%
%or equivalently
%%
%\begin{equation}
%\frac{\Delta F}{F_0} = k^2 \frac{L_D(u_{10},u_{20},\mu_p)}{F_0( u_{1eff}, u_{2eff})} 
%\end{equation}
%%
where $\Delta F$ is the flux decrease due to the planet transit, $F_0$ is the
total emitted flux of the star in our direction,  $u_{10}$ and $u_{20}$ are the
limb darkening coefficients of the unmaculated stellar surface and index $p$
stands for the planet.

If we have one circular Type I spot on the stellar surface, then this can be
modelled by replacing the stellar flux with the spot's flux at the place of the
spot. Then, we use the new $F_{star}$ value instead of $F_0$ in Eq. (3). In this case,
the total flux of a spherical star with only one Type I spot can be written as
(if the spot is small enough) 
\begin{eqnarray}
F_{star} = &&  \pi R_{star}^2 I_0 ( 1 - u_{10}/3 - u_{20}/6 ) \\ \nonumber 
           && -\pi R_{s}^2 [ I_0 L_{D}(u_{10},u_{20},\mu_s) - I_s
	   L_{D}(u_{1s},u_{2s},\mu_s) ]. \\
	   \nonumber 
\end{eqnarray}
Index $s$ stands for the spotted area. The first term on the right hand side describes the
total flux stemming from the stellar disc, while the second term describes the
effect of the spot: we remove a certain small part of the stellar flux and replace
it by the spot's flux. The argument of the $L_D$ function is $\mu_{s}$ which we can
calculate easily from the spot's position. This equation also shows that in the
case of a dark spot, $F_{star}$ is smaller and the transit depth becomes larger. For
faculae, which are brighter than the surface, $F_{star}$ become larger and hence
the transit depth will be smaller.

When we model the transit, we replace this real star with a hypothetical star of
an unmaculated surface. Although the transit depth will change due to the
presence of stellar spots, it is easy to show that the radius ratio $k$ remains
unchanged and precisely determinable if one chooses the effective limb darkening
coefficients appropriately. 

We use the following abbreviation (cf. Eq. (4)):
\begin{equation}
A = \frac{\Sigma_{j=1}^{N_s} \pi R_{sj}^2 [ I_0 L_{D}(u_{10},u_{20},\mu_{sj}) - I_{sj}
	   L_{D}(u_{1sj},u_{2sj},\mu_{sj}) ]}{\pi R^2_{star} I_0}
\end{equation}
where $j$ stands for the $j$th spot. Then we have a step-by-step calculation 
of the observable light loss during the transit, 
\begin{eqnarray}
\frac{\Delta F}{F_{star}} = && k^2 \frac{1-u_{10}\delta-u_{20}\delta^2}{1-\frac{u_{10}}{3} - \frac{u_{20}}{6} - A} \\ \nonumber
                          = && k^2\frac{1-\frac{u_{10}}{1-A}\delta-\frac{u_{20}}{1-A}\delta^2}{1-\frac{u_{10}}{3(1-A)}-
\frac{u_{20}}{6(1-A)}} + \alpha ,
\end{eqnarray}
where we abbreviate
\begin{eqnarray}
\alpha = \frac{A k^2}{(1-A) \left( 1 - \frac{u_{10}}{3(1-A)} - \frac{u_{20}}{6(1-A)}\right)}.
\end{eqnarray}
This $\alpha$ is constant during the transit, because it does not
depend on the planet-star distance and we assumed above that the spot is fixed in
space, i.e. the star rotational period is much longer than the transit length.

Finally the basic equation which relates the observed light loss during a
transit, the planet-to-stellar radius ratio, stellar spot effects and limb
darkening to each other is
\begin{equation}
\frac{\Delta F - \alpha F_{star}}{F_{star}} = k^2
\frac{1-u_{1eff}\delta-u_{2eff}\delta^2}{1-\frac{u_{1eff}}{3}-\frac{u_{2eff}}{6}}.
\end{equation}
Comparing the equations of this section to each other we can see that the
newly defined effective limb darkening coefficients are related to the values
valid on the unspotted stellar surface such that:
\begin{equation}
u_{1eff} = \frac{u_{10}}{1-A},~~~u_{2eff} = \frac{u_{20}}{1-A}.
\end{equation}
From this mathematical analysis of transit light curves it is obvious now that
spots acts like a contamination source in the aperture and this is valid during the entire transit. If an unresolved star or other light source (also  called
"contamination source") contributes to the observed flux, then that decreases the
observed transit depth. This extra light source is routinely removed from the
light curve in general. Certain spots acts as a negative $\alpha$. In total,
this kind of contamination-like source can be either positive or negative, but
in the transit fits we did not include negative contamination.

It has been already recognized, that Type I spots can act as contamination sources, but
our result implies also that limb darkening coefficients can also change which
is a new result. In addition to the arbitrary sign of $\alpha$, these latter two
equations justify that contamination should be an additional free parameter in
the transit fit which would enable to remove all the spot-effects which may
remain after the baseline-correction since they might not cause light curve
modulation in and out of transit. It also shows that no detailed spot modelling
is needed to fully understand the transit light curves if no spot-crossing
occurs, even if polar or any other spots are present. The problem of this 
case can be solved in the following simple and robust way: limb darkening
coefficients and contamination factor should be free parameters. Last but not
least it shows that the observed limb darkening coefficients can be highly
different from the theoretically predicted values in the presence of spots, 
even if they do not produce light curve modulations. The uniqueness of the
transit light curve solutions are provided by the fact that the length of the
ingress and egress phases, as well as the full transit length, are also related to
the size ratio of the star and the planet (see e.g. Winn 2010).

This analysis is only valid for small spots. The analysis
can, however, be generalized easily for numerous and larger spots, as we did with the
summation above. It is also easy to extend the analysis to bright spots (i.e.
faculae). If one takes more terms in the expression of the limb
darkening laws into account, then the effective limb darkening coefficients will be
transformed in the same way as the first two terms, and it is then clear how one has to extend the $\alpha$-function with additional limb darkening
terms.

We also note that for this spotted case the same applies as for the
previous unspotted case in Section 4: if the limb darkening coefficients are
fitted, then we can determine their exact values if they fulfil the
S/N requirements. We have to keep in mind, however, that we measure the effective 
limb darkening coefficients in this case and not the ones related to an unmaculated stellar surface.

Recently, combined optical and near infrared observations of transits were
proposed by Ballerini et al. (2012) to minimize the effect of spots on the
transit parameters. It was proven therein that it is possible to derive precise
planetary radii from this kind of multicolour observation, if the temperature
of spots and the theoretical limb darkening coefficients are known. Here we have
showed that the same result can be reached without spot parameters and without the
knowledge of the theoretical limb darkening coefficients (which are practically
unkown today). Our method is not only simpler, but it does not require multicolour
photometry. Both methods have the advantage of being able to provide the
exact and precise planetary parameters in highly active cases when the spots are
not crossed by the planet, and therefore we have limited information about the
stellar surface.

\subsection{Effect of more than one active area}

The direct observations of the Sun's surface, the Doppler-imaging of other
active stars and their photometric spot studies show concordantly that, in general, 
more than one active area is present on the stellar surface at the same time.
The results of the previous section can be generalized easily by simple
summing of the effects of several individual active areas (cf. Eq. (5)).

Studies of the Sun show that the total area of the faculae depends 
on the area of the dark spots. According to Chapman et al. (1997), the following
expression is a good approximation for the Sun
\begin{equation}
\Sigma A_{faculae}\approx 17 \Sigma A_{spot} ,
\end{equation}
but the conversion factor of the total area ratio shows small secular variations during one
solar cyle (Chapman et al. 1997). It is reasonable to assume that other stars
have similar relationships between their total area ratios, although the
conversion factor in Eq. (10) can vary. For instance, Chapman et al. (2011) find that
the conversion factor varies from cycle to cycle, and for solar Cycle 23 it was
42 instead of the 17 that is valid for Cycle 22. More studies are required to
determine this factor and its dependence on the stellar properties. Foukal
(1998) shows that there is an anticorrelation between $A_{spot}$ and
$A_{faculae}$ in the case of the Sun and he argues that this can be extrapolated
to other stars.

The total spot coverage varies over a wide range. The Sun has a spot coverage of
0.1\% only at maximum. Several observations suggest that many of the normal 
stars can have 1-2\% spot coverage, but there are quite a few stars showing
10-60\% spot coverage on its surface (Schrijver \& Zwaan 2004 and references
therein; this coverage refers to the total area of the spots relative to
the whole stellar surface). The most active stars seem to be located in close
binary stars and giants, and they are not so important from the point of view of
the presently known transiting exoplanets. But some exoplanet hosting stars (like CoRoT-2,
CoRoT-6, CoRoT-7 etc.) seem to be very active stars. E.g. Lanza et al. (2009)
established $\sim7$\% spot coverage for CoRoT-2.

The changes in the limb darkening coefficients have dramatic conclusions for the
errors in planet-to-stellar radius ratio if one fixes the limb darkening
coefficients according to tabulated values and neglects the effect of spots on the
limb darkening. If one accepts the conclusions of Jeffries \& Jackson (2012),
namely that $\sim50$\% of the surface can be covered by small spots in an active
star without observable photometric modulation (but spectroscopic activity
indicators still imply an active nature of this hypothetical star), then the
$A$-function defined by Eq. (5) can have a value of $\sim0.36$. This causes via
Eq. (9) the observable limb darkening coefficients to increase by a
factor of 1.56 - if we do not take into account the role of faculae, which may 
be negligible according to the extrapolation suggested by Foukal (1998). If the
host star is less active, we take 5\% spot coverage as an example, the
$A$-function will cause a change of $\sim0.037$ in the limb darkening
coefficients, which will lead to 9\% error in the radius ratio (cf. Eq. 1)). If the
star has only 0.5\% spot coverage, then the $A$-function will cause a change of
only 0.004 in the limb darkening coefficients, but this still causes 1\% error in
the radius ratio. Thus, 0.5\% spot coverage can be considered as a limit: below
it one can neglect the change in the limb darkening coefficients owing to
Type I stellar spots. These changes in the observed effective and the tabulated limb darkening coefficients are plotted in Figure 5. 
(For this estimation we used a rather grazing eclipse. For central eclipses the effect is less but still considerable.)

We note that when a host star of a transiting planet is highly active, like
CoRoT-2, then it is not enough to take an appropriate contamination
factor into account to change the transit depth or to carefully model the crossing of Type II
spots, including a baseline correction to remove all the contributions of
spots to the photometry. It is also necessary to keep the limb darkening
coefficients as free parameters or to modify them according to Eq. (9), because
the modification of the limb darkening coefficients will also cause non-negligible effect on the planetary radius.

Thus we conclude that stellar activity modifies both the observable effective limb
darkening coefficients, as well as the observable transit depth. Then, if spots and faculae are present, one cannot fix
the limb darkening coefficients corresponding to the stellar effective surface
temperature. It is preferable to leave them as adjustable parameters,
because it is difficult to predict the effect of the spots on the limb darkening
coefficients. We also established that the effective limb darkening coefficients (note that almost all transit light curve modellers use these) 
can have unphysical values because of spots and faculae. Of course, 
`unphysical' is only the correct word if, and only if, we do not compare the spotted
star to an unspotted model calculation. It is also clear that in such a case,
we also can determine with good accuracy ($<1$\%) the radius ratio of the star and
the planet, the impact parameter, and the $a/R_s$ ratio and the limb darkening
coefficients, but we just keep in mind, that in the case of several spots and
faculae, the limb darkening coefficients yielded by the modelling do not correspond to the limb
darkening coefficients of an unmaculated star with the same effective
temperature, but instead correspond to the effective limb darkening
coefficients.

%
%                                                One column figure
%----------------------------------------------------------- refl
   \begin{figure}
   \centering
   \includegraphics[width=9cm]{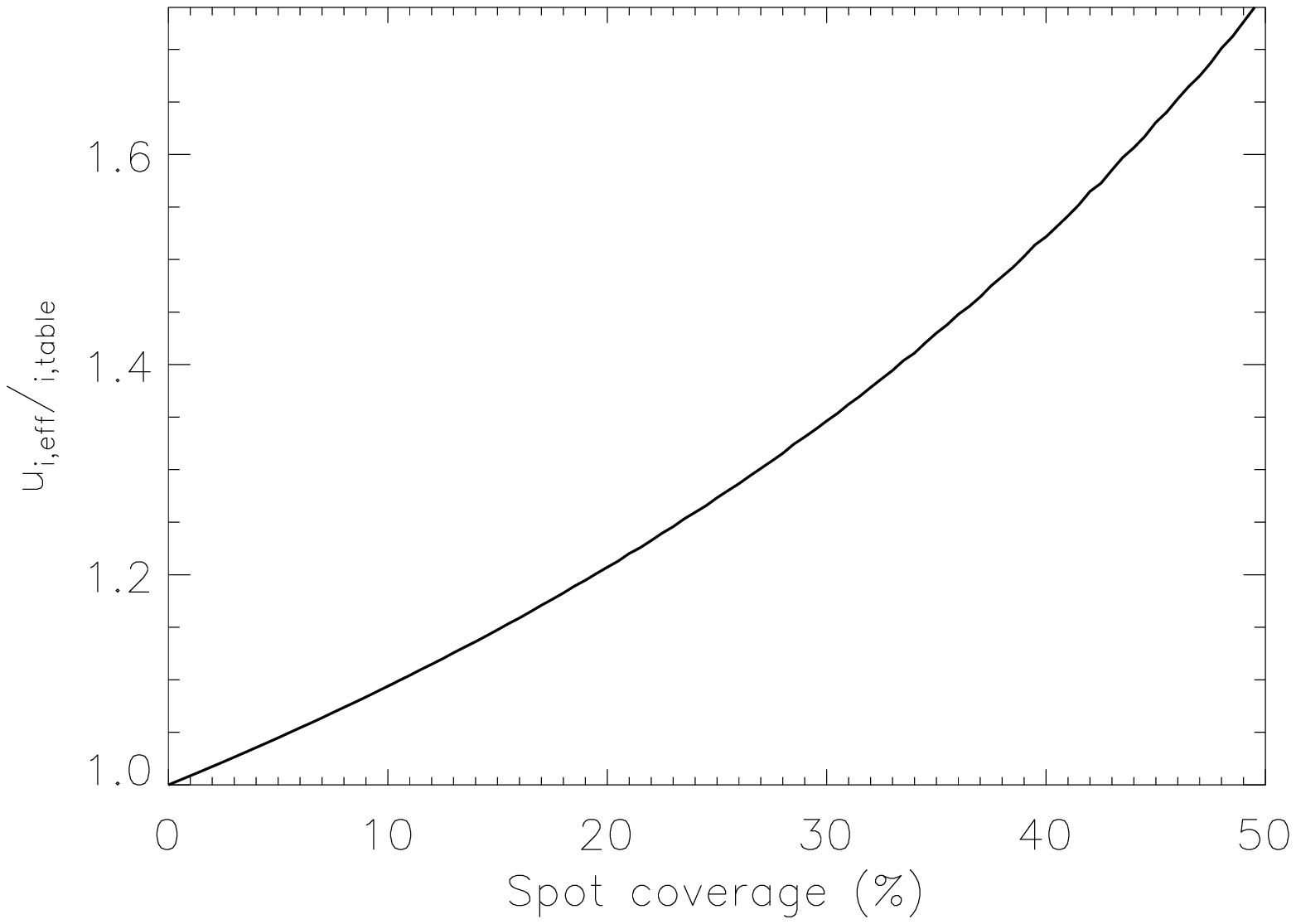}
      \caption{The x-axis is the total spotted area in percentage of the whole 
       stellar surface area. The y-axis is the effective - i.e observed - limb 
       darkening coefficients relative to the table value at the given stellar 
       efffective temperature. For this figure we used $T_{star}=5775$K and 
       $T_{spots}=3775$K, and the positions of the spots were chosen randomly 
       on the visible hemisphere. The size of the spots were always the same, 
       so higher spot coverage corresponds to larger number of spots. The limb 
       darkening coefficients were taken from Claret \& Bloemen (2011) for 
       $R$-band, $\log g = 4.5$ and at solar metallicity at these  
       temperatures. Effects of faculae were not considered in this figure.} 
         \label{contours}
   \end{figure}
%
%______________________________________________________________

\section{Determinability of individual limb darkening coefficients}

Here we study the determinability of the specific limb darkening
coefficients. Since many combinations of limb darkening coefficients are able to
reproduce the same surface brightness distribution within the errors of the
observations, it seems likely that the light curve fitting procedures
reproduce the intensity distribution and not the true values of the
coefficients. In other words, the inversions of the polynomial expressions of
the limb darkening laws incorporate many degeneracies. That is why we are able
to determine all the transit parameters with good accuracy and to reproduce the
corresponding stellar surface brightness distribution within the errors of the
observations, but we are not able to determine the individual values of the limb
darkening coefficients. From the point of view of exoplanet parameter
determination this does not present any difficulties. To illustrate this, we
rewrite Eq. (1) for the following analysis as 
\begin{equation}
k^2 = \frac{\Delta F / F}{L_D}.
\end{equation}
The current aim is to know the planetary radii with 5\% precision (Valencia et al. 2007), although the future goal is 1\% (Wagner et al. 2011). To determine $k$, say, with $5$\% precision, it is enough to know the
limb darkening with a precision of 10\%, which follows from the linear
error analysis
\begin{equation}
\frac{\Delta k}{k} = \frac{1}{2} \left(\frac{\Delta F}{F} + \frac{\Delta L_D}{L_D} \right),
\end{equation}
because the photometric error is practically negligible for space-based
photometry beside the error steming from limb darkening's errors (cf. Section
3). This means that the transit light curve modelling codes can produce quite
different limb darkening coefficients than what we expect. Now we aim to
determine these differences.

%
%                                                One column figure
%----------------------------------------------------------- refl
   \begin{figure}
   \centering
   \includegraphics[width=9cm]{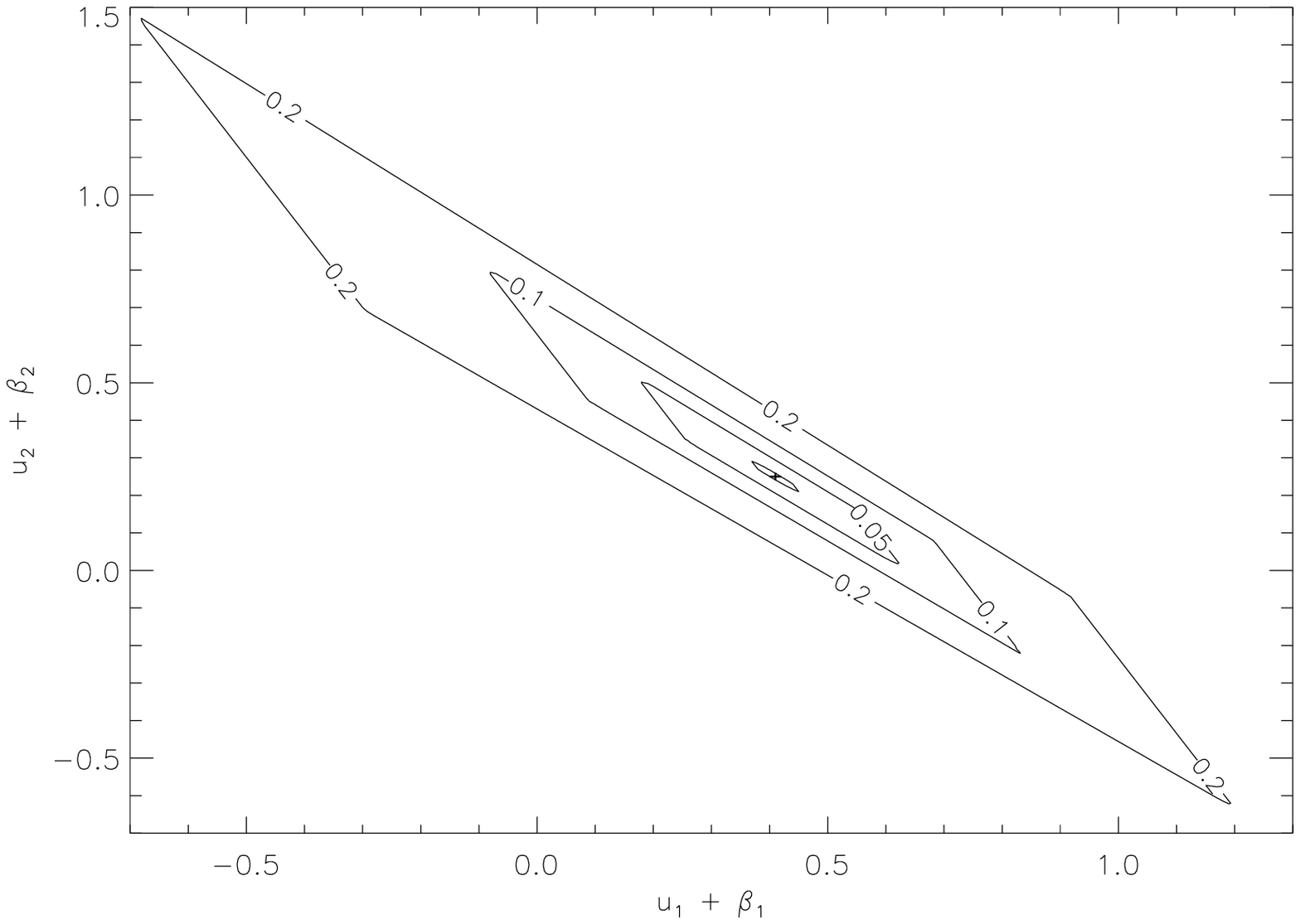}
      \caption{The x-axis is the modified linear limb darkening coefficient $x_1 = u_1
      + \beta_1$, the y-axis is the modified quadratic limb darkening
      coefficients $x_2 = u_2 + \beta_2$. $u_1$, $u_2$ are fixed. The point in  
      the middle shows the $u_1 = 0.41$, $u_2=0.25$ point, i.e. the original 
      value of the limb darkening coefficients of the example presented in 
      Section 8. Contours represent the quantity $\epsilon_{\mathrm{max}}$, i.e.
      the absolute value of the maximum deviation of the original limb darkening 
      function with $u_{1,2}$ coefficients from the limb darkening function with
      $x_{1,2}$ ($\delta$ run between 0 and 1).}
         \label{radialintensityprofiles}
   \end{figure}
%
%______________________________________________________________

%
%                                                One column figure
%----------------------------------------------------------- refl
   \begin{figure}
   \centering
   \includegraphics[width=9cm]{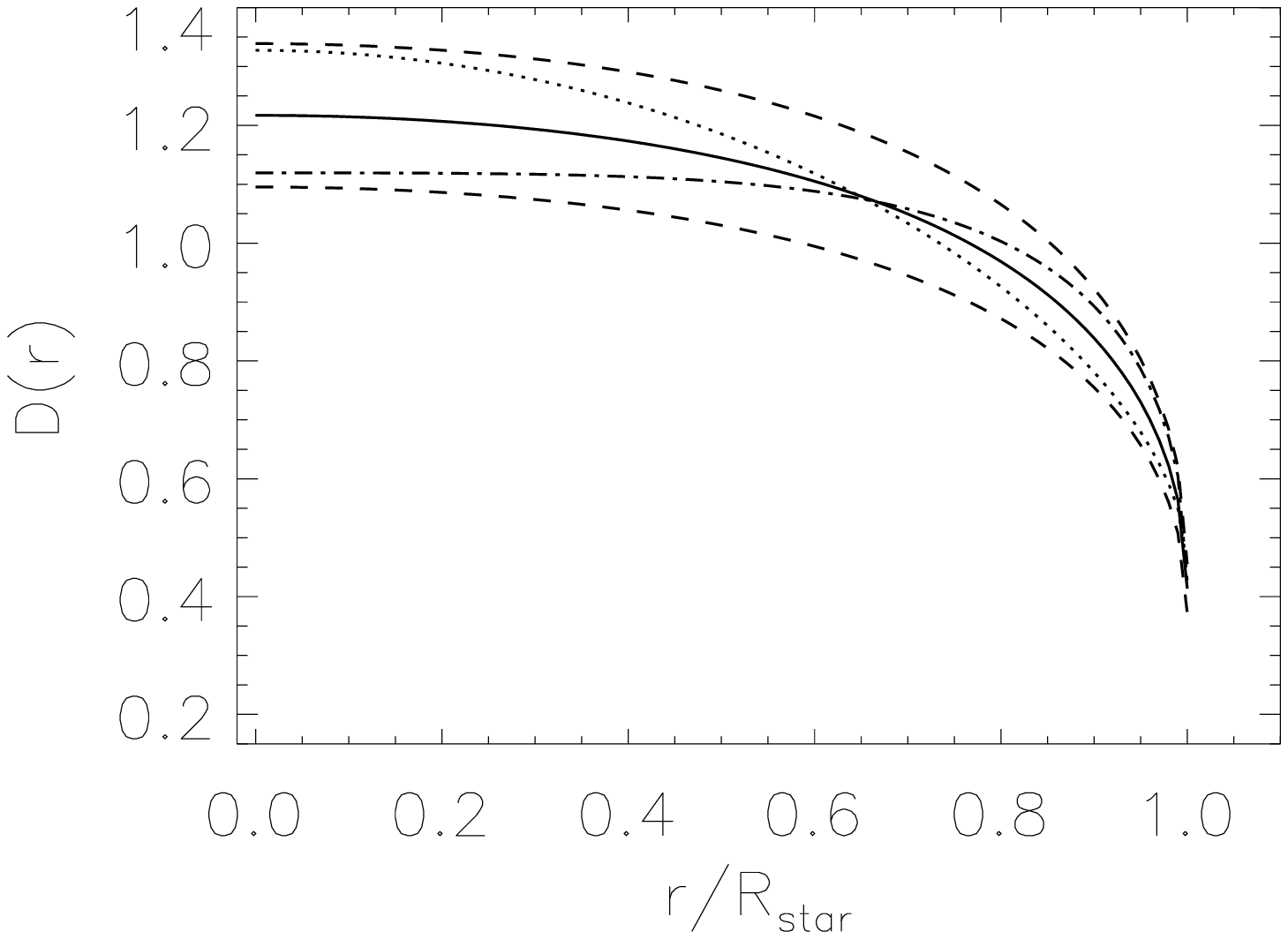}
      \caption{The radial intensity profile $D(r/R)$ as a function of the
      sky-projected distance $r$ from the stellar centre ($R$ is the stellar
      radius). As in all calculation in this paper, $\mu = \sqrt{1-(r/R)^2}$. 
      The solid line shows the effect of limb darkening $D(r) = L_D(u_1, u_2,
      \mu)/(1-u_1/3 -u_2/6)$ that - multiplied by $k^2$ - is directly 
      proportional to the light loss during a transit. The dashed lines show
      the tolerable ranges: between these lines the radial intensity
      distribution profile will produce a radius ratio $k$ that is in the
       tolerance range of $\pm5$\%. The dotted line is an example of
      an acceptable radial intensity distribution profile with $u_1=0.82,
      u_2=-0.16$, while the dot-dashed line is with $u_1=0.02$, $u_2=0.6$. }
         \label{contours}
   \end{figure}
%
%______________________________________________________________

We denote the expected limb darkening coefficients by $u_i$, and the
difference between the measured and the expected values by $\beta_i$. The
relative deviation between measured and expected stellar  intensity profiles is
denoted by $\epsilon(\delta)$ and the maximum allowed relative deviation - occurs at the still unknown $\delta$ - is denoted by $\epsilon_{\mathrm{max}}$, so we
can write 
\begin{equation}
L_D(u_1, u_2, ..., \delta) - L_D(u_1+\beta_1, u_2+\beta_2, ..., \delta) = \epsilon_{\mathrm{max}}.
\end{equation}
Although the analysis can easily be extended to higher orders, we cut it at
second order because this is generally used for transit analysis (but as
 can be seen, the conclusion does not change significantly if one extends the
calculation to higher orders). 

We chose a solar-like star to illustrate the operation of the
$\beta_i$-numbers. We plotted the deviation $|\epsilon|$ for different cases in
Figure 6. In that figure, there is a rather wide range for $\beta_i$
parameters that produce the same limb darkening function within $\pm \epsilon$
tolerance range. It is easy to construct an $L_D$ function, for example, 
of limb darkening coefficients of $u_1+\beta_1 = 0.8$ and $u_2+\beta_2=-0.1$ and
this produces the same results within the tolerance range as the
correct values $u_1=0.41$ and $u_2=0.26$. This example is presented in Figure 6. Along 
the sides of the  paralellograms one can easily construct the same quality of the light
curve fit (cf. Eq. (13). In Figure 7 we plotted another example: the stellar surface brightness
distribution for different $\beta_i$ numbers. As one can see there is a high
tolerance range, so one can choose very different limb darkening combinations
from the theoretical predictions that produce a similar surface brightness
distribution within the tolerance range. This example shows that we are able to 
produce acceptable $D(r/R=\mu)$ functions within the tolerance ranges with quite 
different limb darkening coefficients. If the tolerance range changes due to different 
precision requirements, the acceptable range of the limb darkening coefficients will also change.

One can envisage that a transit light curve modelling code may take different numbers as input 
for numerical reasons or owing to data points/error distributions than
the theoretically expected values. This can be also the case if stellar spots
modify the limb darkening coefficients. That is why the absolute values of
the limb darkening coefficients are not important but the shape of the
radial intensity distribution that they define is (see Figure 7 for such a
comparison).

This analysis gives hints and arguments that what we derive from the transit
light curve are not the true individual values of the limb darkening coefficients
but some rather complicated combination of them. The limits of the potentially  wide ranges depend on the actual impact parameter
and photometric precision (i.e. the range of the tolerance factor
$\epsilon$), so that to give the limits requires case studies.

From Figure 6 it is also clear that the fitted limb darkening coefficients are
also correlated with each other. This behaviour has already been observed by numerical
experiences in Brown et al. (2001) and in e.g. Deleuil et al. (2008) and Csizmadia
et al. (2011). Brown et al. (2001) and P\'al (2008) suggest fitting the
combination $u_+ = u_1 + u_2$, $u_- = u_1 - u_2$ and $u_+ = u_1 \cos 40^\circ
+ u_2 \sin 40^\circ$, $u_- = u_1 \cos 40^\circ - u_2 \sin 40^\circ$ instead
of fit $u_1$ and $u_2$ themselves because they are less correlated to each other
as previously assumed. As is clear from Figure 6 this rotational-like
transformation can help better constrain one of these combinations, but the
other will still feature a value in a wide range and the problem cannot be surpassed in this way. The
values that look `unphysical' at first sight may describe the radial intensity distribution as well as do other values. We emphasize again that the absolute  values of the fitted limb
darkening coefficients are themselves not important but the radial intensity
profiles are. These profiles, taking  the tolerable deviations into account, can be
produced with a very wide ranges of limb darkening coefficients. That is why the
basis of the analysis of the results are not comparison of the resulting  limb
darkening coefficient values themselves to some theoretical table values, but the comparison of the resulting radial intensity profiles of the
predicted with the observed ones, a diagram similar to Figure 7.

\section{Conclusions}

%We summarize the results of this paper as follows.

In this paper we have attempted to understand the impact of limb darkening on
the planetary radius, to present ideas as to which effects can modify the
theoretically predicted limb darkening coefficients, and to asnwer why some authors have
highly different observed limb darkening coefficients from the theoretical
predictions.

%In Sections 1, 2 and 3 we gave an introduction to the problem.

We concluded that concurrent limb darkening tables do not yield 
consistent results in several temperature- and metallicity regions, and we
showed that the inconsistencies will lead to serious defects in the
determination of the planet-to-stellar radius ratios if one fixes the limb
darkening coefficients according to these tables. We also proved with numerical 
experiments that one can safely fit the limb darkening coefficients.

We investigated how the presence of stellar spots - both dark
spots and bright faculae - modify the observable transit depth and the limb
darkening coefficients. We found that below 0.5\% spot coverage, the change in
the effective limb darkening coefficients causes no significant error in the
measured planet-to-stellar radius ratio, but over that coverage this cannot be neglected, 
and neglecting the change in the limb darkening coefficients themselves has a significant 
change in the radius ratio.

We pointed out that the radial intensity profiles of the stellar
discs are not very sensitive to the specific values of the limb darkening
coefficients, and they are anticorrelated to each other in the case of the
quadratic limb darkening law (cf. Figure 6), and a similar statement can be made  about
higher order limb darkening laws. Consequently, the observed values of the limb
darkening coefficients can be very different from the theoretically tabulated
ones even if they produce the same radial intensity profiles within the
tolerance.

Limb darkening is still a problem for transit modelling. On the one hand, limb
darkening is a challange for stellar astrophysics. %, because we do not understand
%well the stellar atmospheres if the limb darkening values are not measurable
%with sufficient accuracy. 
On the other hand, we can replace the limb darkening coefficients of an
unmaculated star with the appropriately weighted effective limb darkening
coefficients for the modelling %, and then the modeling will give satisfactorically
%results with the required precision, if the signal-to-noise ratio is high
%enough. 
Our final conclusion is that the transit parameters can be determined even if
our knowledge of limb darkening and stellar spots is not satisfactory. Our work
will make it clear that one has to fit the limb darkening coefficients and it is
not self-evident that the resulting limb darkening coefficients should 
agree with the theoretical predictions. Indeed, these can be very 
different. Our results suggest not only that fitting the limb darkening coefficients 
will produce more reliable solutions but also that their adjustment will not
affect the accuracy of determining other transit parameters. 
However, predicting the effective limb darkening coefficients is
difficult, because it requires a reliable and detailed spot solution. Moreover,
these can vary from transit to transit as the spots evolve. If we fit them, then
we can determine their values, too, but then such effective limb darkening
coefficients have no direct link to the theoretical limb darkening tables. 
When a statistically significant sample of high-quality, multicolour
photometric analyses of transits become available, then the comparison of the
resulting limb darkening coefficient to the available tables will 
become useful to characterize the discrepancies between theory and
observations better. This can contribute significantly to improving our knowledge of 
stellar atmospheres. It is also important to simultaneously determine the
activity level of the star and the limb darkening coefficients%, because as we
%have pointed out the effective (i.e. observable) limb darkening
%coefficients depend on the actual spot-coverage and activity level of the star.
%We recall that stellar spots and faculae are time-variable phenomena.

It is also clear that further detailed studies are required to completely understand
 the limb darkening phenomenon and its interaction with e.g. gravity
darkening. Such studies are very useful for planning such projects as PLATO or
for understanding the present-day ground- and space-based measurements more deeply.

%
%____________________________________________________________________________
\begin{acknowledgements}
This research has made use of the SIMBAD database, operated at the CDS, Strassbourg,
France. We thank an anonymous referee for his/her comments.
\end{acknowledgements}

%
%________________________________________________________________
%\bibliographystyle{bibtex/aa}

%\bibliography{bibl}

\begin{appendix}

\section{The limb darkening laws}

Limb darkening is a stellar atmosphere phenomenon that is reviewed in detail
in several textbooks (see e.g. Kallrath \& Milone 2009). The most general
formula to describe the effect of limb darkening is
\begin{eqnarray}
I(\gamma) = && I_0 L_D (u_1,~u_2,~...,~u_i,~..., \mu) = \\  \nonumber
            &&I_0 \left(1 -  \Sigma_{j=1}^{\infty}u_j (1-\cos\gamma)^j \right), \\ \nonumber
\end{eqnarray}
where $u_j$ are the so-called limb darkening coefficients, $\gamma$  the angle 
between the normal vector of the surface point and the direction to the observer, 
$I_0$ and $I(\gamma)$ are the surface brightnesses (or intensities) at the appearant 
centre of the stellar disc and in a certain point of the stellar surface characterized 
by the angle $\gamma$, and $L_D$ is the limb darkening function defined by Eq.
(A.1). We also use the common notation $\mu = \cos \gamma$.

Many studies of eclipsing binaries have indicated that at least two terms in Eq. (A.1) 
should be kept to fit the light curves satisfactorily (Twigg \& Raffert 1980;
Alencar \& Vaz 1999; Albrow et al. 2001; Claret 2000, 2008, Southworth et al.
2007). That is why, in the exoplanetary research field it is widely accepted to use the truncated
formula of Eq. (A.1)
\begin{equation}
L_D = 1-\Sigma_{j=1}^{2}u_j (1-\cos\gamma)^j ,
\end{equation}
which can easily be rewritten as
\begin{equation}
L_D = 1- (u_1 + u_2) + (u_1 + 2 u_2) \mu - u_2 \mu^2 .
\end{equation}
There are different approaches  describing the limb darkening
effects, including logarithmic or square root terms of $\mu$, see e.g. Claret (2000) or van Hamme (1993) for an overview.

\section{Transit depth and limb darkening}

It is straightforward to show in the small-planet approximation that the transit depth can
be written as
\begin{equation}
\frac{\Delta F}{F} = k^2 \frac{1-u_1 (1-\mu) - u_2 (1-\mu)^2}{1-\frac{u_1}{3} - \frac{u_2}{6}} ,
\end{equation}
where
\begin{equation}
\Delta F = \pi R_{planet}^2 \left({1-u_1 (1-\mu) - u_2 (1-\mu)^2}\right) \nonumber
\end{equation}
and
\begin{equation}
F = \pi R_{star}^2 \left({1-\frac{u_1}{3} - \frac{u_2}{6}}\right) \nonumber
\end{equation}
also hold, and it is trivial to extend these expressions to higher orders of the
limb darkening approximations (cf. Kallrath \& Milone 2009 and Mandel \& Agol
2002). In these expressions $\Delta F$ is the light lost due to the transit; $F$
is the stellar flux out of the transit part of the light curve; $k=R_{planet} /
R_{star}$ is the ratio of the planet radius $R_{planet}$ and the star's radius
$R_{star}$.  In addition, in the moment of the maximum light lost, $\mu = \sqrt{1
- b^2}$ where $b$ is the impact parameter. If we are interested in the light
loss at any arbitrary time inside the transit, then $\mu = \sqrt{1 - \delta^2}$
where $\delta$ is the sky-projected distance measured in stellar units. (For details on how to
determine it for any arbitrary time see Kallrath \& Milone 2009.)

Taking the typical values of the limb darkening coefficients into account (see
Section 2 or the references mentioned in Section 1), one can easily see that
limb darkening will modify the true depth by up to a factor of $\sim2$ relative
to the case where one takes the transit depth to be proportional only to the
radius ratio - therefore the effect of limb darkening is not negligible.

\end{appendix}

\end{document}